\documentclass[a4paper,11pt]{article}
\pdfoutput=1 

\usepackage{jheppub} 
\usepackage{overpic}
\usepackage[T1]{fontenc} 

\title{Thermal properties of light mesons from holography}

\author[a]{Xuanmin Cao,}
\author[a]{Songyu Qiu,}
\author[a]{Hui Liu}
\author[a,1]{and Danning Li\note{Corresponding author.}}

\affiliation[a]{Department of physics and Siyuan Laboratory, Jinan University, Guangzhou 510632, China}

\emailAdd{caoxm@jnu.edu.cn}
\emailAdd{songyuqiu@stu2018.jnu.edu.cn}
\emailAdd{tliuhui@jnu.edu.cn}
\emailAdd{lidanning@jnu.edu.cn}

\abstract{The thermal properties of light mesons, including the temperature dependence of their masses (both screening and pole masses) and thermal widths, are studied in a two-flavor ($N_f=2$) soft-wall AdS/QCD model. By solving the spatial correlation functions, we extract the screening masses ($m_{\rm{scr}}$) from their poles. The screening masses of pseudo-scalar ($\pi$) and axial-vector ($a_1$) mesons increase almost monotonously with the increase of temperature. The screening masses of scalar ($\sigma$) and vector ($\rho$) mesons decrease at low temperature and increase at high temperature. The pole masses ($m_{\rm{pole}}$) and the thermal widths ($\Gamma$) are extracted from the temporal correlation functions and the corresponding spectral functions. The results indicate that the pole masses have local minima at low temperature and increase at high temperature. The thermal widths increase rapidly above the chiral crossover temperature $T_{cp}$, indicating the dissociations of mesons at high temperature. Furthermore, the degeneration of the chiral partners ($\pi$ and $\sigma$, $\rho$ and $a_1$) above $T_{cp}$ is observed from the screening and pole masses, revealing the chiral symmetry restoration at the hadronic spectrum level. Finally, we numerically verify that the spectral functions in the temporal regime are strongly related to the quasi-normal modes with complex frequencies $\omega_0=m_{\rm{pole}}-i \Gamma/2$.}

\begin{document}
\maketitle
\flushbottom
\section{Introduction}
\label{sec:intro}
It is widely accepted that there is a crossover transition from the normal hadronic phase to the quark-gluon plasma (QGP) phase for Quantum Chromodynamics (QCD) matter at a temperature around $150-170$ MeV~\cite{Pisarski:1983ms,deForcrand:2006pv,Ding:2015ona}. It is one of the main goals of heavy ion collisions to probe this transition~\cite{Adams:2005dq}. To get a better understanding of the experimental data, it is essential to acquire a complete knowledge of the in-medium properties of hadrons, especially of light mesons.

In vacuum, mesons are described by their masses and decay widths in different channels. Due to the breaking of Lorentz symmetry at finite temperature, the real part of the dispersion relation might be modified as $\omega^2(\bold{p})=u^2\bold{p}^2+m^2$~\cite{Shuryak:1990ie}, where $u$ is the velocity. One can define two kinds of masses, i.e., the screening mass and the pole mass~\cite{Shuryak:1990ie,DeTar:1987ar,DeTar:1987xb,Son:2001ff,Son:2002ci}. Both of them carry information about the correlation function of the meson field in the hot medium.

The screening mass ($m_{\rm{scr}}$) characterizes the color screening effect and determines an important length scale of nuclear force inside the hot medium~\cite{DeTar:1987ar,DeTar:1987xb}. Quantitatively, it describes the exponential decay of the long range spatial correlation function of the meson field, i.e., $\tilde{G}(\bold{x})\sim e^{-m_{\rm{scr}} |\bold{x}|}/ |\bold{x}|$ when $|\bold{x}|>>m^{-1}_{\rm{scr}}$.  In momentum space, it is the pole of the spatial meson correlation function~\cite{Schmidt:1999je,Son:2001ff,Aarts:2017rrl,Aarts:2018glk}, i.e., $ G^{-1}(\bold{p})|_{\bold{p}^2=-m^2_{\rm{scr}}}=0$. Furthermore, as shown in Refs.~\cite{Cheng:2010fe,Bazavov:2019www},  it might also carry information about the restoration of the chiral symmetry and the $\rm{U_A(1)}$ symmetry in the hot medium.

Unlike the screening mass, the pole mass ($m$) characterizes the temporal correlation function ($G(\omega)$ or $\tilde{G}(\tau)$). It can be defined as the real part of the pole of the temporal correlation function~\cite{Son:2001ff,Ayala:2012ch}, or equivalently the resonance peak location of the spectral function~\cite{Nakahara:1999vy,Asakawa:2000tr}. The dispersion relation makes it easy to get the simple connection between these two kinds of masses, i.e., $m=m_{\rm{scr}} u$. In vacuum, $u=c=1$ (with $c$ the speed of light),  and one has $m=m_{\rm{scr}}$. Due to the effective interaction between the mesons and the medium at a finite temperature $T$, both kinds of masses would be modified~\cite{Li:1995qm,Rapp:1995zy,Adamova:2006nu,Arnaldi:2006jq,Rapp:2009yu,Aarts:2017rrl,Aarts:2018glk}. Generally, the velocity $u(T)$ is less than $1$, so the pole mass would be smaller than the corresponding screening mass~\cite{Pisarski:1996mt}.

Besides these two kinds of masses, the hadronic width is another important quantity to describe hadrons in the medium. It is also a temperature-dependent quantity. The variations of the masses and the effective couplings would change the decay width of the meson. Moreover, the scattering channel would modify the interaction rate, and even a stable hadron would develop a thermal width~\cite{Ayala:2012ch}.  The rapid and monotonic increase of the thermal width is a possible signal of a phase transition~\cite{Ayala:2012ch,Chen:2020afc}. When the thermal width increases to a sufficiently large value, the meson is dissociated. In this situation, the meson correlation function is still meaningful~\cite{Mocsy:2005qw}. The masses and the width still contain specific information about the correlation function. These are widely considered in the lattice simulations, which give results up to several times the chiral crossover (pseudo-critical)  temperature $T_{cp}$~\cite{Cheng:2010fe,Bazavov:2019www}.

So far, many efforts have been made to study the in-medium hadronic properties, including the lattice QCD (LQCD) simulations~\cite{Brandt:2015sxa,Brandt:2014qqa,Cheng:2010fe,Bazavov:2019www,
Aarts:2017rrl,Aarts:2018glk,Ding:2017std}, the chiral perturbation theory ($\chi$PT)~\cite{Son:2001ff,Son:2002ci}, the Nambu-Jona-Lasinio models (NJL)~\cite{Hansen:2006ee,Jiang:2011aw, Ebert:1992jx,Sheng:2020hge}, the functional renormalization group (FRG)~\cite{Tripolt:2013jra,Wang:2017vis}, the Dyson-Schwinger Equation (DSE) and the Bethe-Salpeter Equation (BSE)~\cite{Fischer:2018sdj,Horvatic:2010md,Chen:2020afc,Gao:2020hwo}.
Consistent results are given by the different methods that the meson masses increase with the increasing temperature when the temperature is above the chiral crossover temperature $T_{cp}$. At extremely high temperature, the screening mass would increase linearly with $T$~\cite{Florkowski:1993bq}. But, there is no firm consensus on the behavior at the temperature below or around $T_c$ due to the tricky strong coupling problem. For example, the NJL and DSE studies~\cite{Hansen:2006ee,Fischer:2018sdj} give monotonically increasing pion pole mass. Nevertheless, the $\chi$PT analysis~\cite{Son:2001ff,Son:2002ci}, the LQCD simulations~\cite{Brandt:2014qqa,Brandt:2015sxa}, and the NJL model with gluon condensate~\cite{Ebert:1992jx} indicate that the pion pole mass decreases with the increasing temperature when the temperature is below $T_{cp}$. Therefore, it is still meaningful to try to obtain more information from other nonperturbative methods.

The holographic method developed from the AdS/CFT correspondence~\cite{Maldacena:1997re,Gubser:1998bc,Witten:1998qj,Kovtun:2004de} does provide another powerful tool for dealing with the strong coupling QCD. There are many well-constructed holographic QCD models in the bottom-up approach, like the hard wall model~\cite{Erlich:2005qh}, the soft-wall model~\cite{Karch:2006pv}, the Einstein-Maxwell-Dilaton systems~\cite{Gubser:2008ny,Gubser:2008yx,DeWolfe:2010he,Gursoy:2007cb,Gursoy:2007er} and the light-front holographic QCD~\cite{Brodsky:2014yha}. Among these models, the soft-wall AdS/QCD model and its extensions can describe the hadronic spectrum and relevant quantities~\cite{Gherghetta:2009ac,Kelley:2010mu,Li:2012ay,
Li:2013oda,Sui:2009xe,Colangelo:2008us,Ballon-Bayona:2020qpq,FolcoCapossoli:2019imm}, and the chiral phase transition~\cite{Colangelo:2011sr,Chelabi:2015cwn,Chelabi:2015gpc,Fang:2015ytf,Li:2016gfn,Li:2016smq,
Bartz:2016ufc,Fang:2016nfj,Bartz:2017jku,Fang:2018vkp,Cao:2020ske}. It provides an excellent scenario to consider the light scalar ($\sigma$), pseudo-scalar ($\pi$), vector ($\rho$), axial-vector ($a_1$) mesons simultaneously in a consistent way. Therefore, in this paper, we will study the thermal properties of light mesons in this model.

The holographic framework is widely applied to extract the pole mass and thermal width from the corresponding spectral function~\cite{Erdmenger:2007ja,Erdmenger:2008yj,Kaminski:2009ce,Colangelo:2009ra,Fujita:2009ca,Cui:2014oba,Braga:2016wkm,Vega:2017dbt,Zollner:2020nnt,MartinContreras:2021bis}, which are calculated from the imaginary part of the retarded correlation function. However, as noted in Ref.~\cite{Colangelo:2009ra}, at high temperature, the thermal width increases rapidly, and it is hard to determine the exact location of the resonance peak. A more straightforward method, which maps the pole mass and thermal width to the real and imaginary parts of the complex frequency of the corresponding quasi-normal mode (QNM)~\cite{Frolov:1998wf,Kokkotas:1999bd}, was discussed in Ref.~\cite{Miranda:2009uw} and applied in holographic QCD~\cite{Grigoryan:2010pj,Mamani:2013ssa,Mamani:2018uxf,Braga:2019yeh}. There are also discussions about Debye screening masses extracted from the spatial correlation functions of $CT$-odd operator $\text{Tr}F_{\mu\nu}\tilde{F}^{\mu\nu}$~\cite{Bak:2007fk,Finazzo:2014zga}, the pseudo-scalar glueballs~\cite{Braga:2017apr} and the Polyakov loops~\cite{Andreev:2016hxm}. The Debye screening mass is shown to increase linearly with $T$~\cite{Finazzo:2014zga}, which is consistent with the 4D studies. However, the studies for the light mesons (like the pion and the $\sigma$ meson), and their relationships with phase transitions, are still quite limited in holographic approaches.

Thus, it is very interesting to investigate the thermal properties of the light mesons in the holographic framework, not only for their temporal correlation functions and dissociations but also for their spatial correlation functions and the relations with phase transitions. In our previous work~\cite{Cao:2020ryx}, the thermal pole masses of the pion and the $\sigma$ meson have been extracted in an IR-Improved AdS/QCD model through the spectral function method. The property of the decreasing pion pole mass is qualitatively consistent with Son and Stephanov's general analysis in Refs.~\cite{Son:2001ff,Son:2002ci}, the LQCD simulations in Refs.~\cite{Brandt:2014qqa,Brandt:2015sxa}, and the NJL result in Ref.~\cite{Ebert:1992jx}. However, since the fast broadening of the resonance peak, it is hard to extract the exact thermal width. Here, we will extend our previous study and follow Ref.~\cite{Miranda:2009uw} to investigate the thermal width. It would also be interesting to extend our analysis to the spatial correlation functions and study the screening masses of the light mesons. A complete analysis including the two pairs of chiral partners, $(\pi,\sigma)$ and $(\rho, a_1)$, would be important for understanding the relationship between the hadronic spectra and the chiral phase transition.

The rest parts of this paper are organized as follows. In Sec.~\ref{II}, we will briefly review the soft-wall AdS/QCD model and the chiral phase transition. In Sec.~\ref{III}, we will consider the spatial correlation function and extract the temperature-dependent screening mass in the chiral limit and with the physical quark mass. Then, we will turn to the temporal correlation function and extract the thermal pole mass and the width by solving the QNM frequency in Sec.~\ref{IV}. Finally, in Sec.~\ref{V}, we will give a summary and discussion.

\section{A brief review of the soft-wall AdS/QCD model}\label{II}
The $N_f=2$ soft-wall AdS/QCD model is constructed with the $\rm{SU_L(2)\times SU_R(2)}$ gauge symmetry under the dual 5D geometry~\cite{Karch:2006pv}.
It extends the hard-wall AdS/QCD model~\cite{Erlich:2005qh} by replacing the hard cutoff with a quadratic dilaton field $\Phi(z)=\mu_g^2 z^2$, which depends on the fifth dimension $z$. The action takes the following form
\begin{eqnarray}\label{action}
    S=\int d^5 x \sqrt{g} e^{-\Phi(z)}{\rm{Tr}}\left\{ |D_MX|^2-V(|X|)-{1\over 4g_5^2}\left (F^L_{MN}F^{MN,L}+F^R_{MN}F^{MN,R}\right )\right\},
\end{eqnarray}
where $\sqrt{g}$ is the determinant of the metric. The gauge coupling constant $g_5$ is equal to $2\pi$ when the number of colors $N_c$ is equal to $3$ ~\cite{Erlich:2005qh,Son:2003et}.
$X$ is a matrix-valued scalar field, and its covariant derivative is defined as
\begin{equation}
D_M X=\partial_M X-iL_MX+iXR_M,
\end{equation}
with $L_M$ and $L_R$ the chiral gauge fields,
\begin{equation}
L_M=L_M^i t^i,\qquad R_M=R_M^i t^i.
\end{equation}
The $\rm{SU(2)}$ generators are $t^i=\sigma^i/2$ ($i=1,2,3$) with $\sigma^i$ the Pauli matrices. The potential term is
\begin{equation}
V (|X|) = m_5^2|X|^2 + \lambda |X|^4,
\end{equation}
with $m_5^2$ the 5D mass (taking the value $-3$ in the original soft-wall model)  and $\lambda$ a free parameter. $F_{MN}^{L/R}$ are the field strength tensors of the corresponding chiral gauge fields, which are defined as
\begin{subequations}
\begin{eqnarray}
F_{MN}^{L}=\partial_ML_N-\partial_NL_M-i[L_M,L_N],\\
F_{MN}^{R}=\partial_MR_N-\partial_NR_M-i[R_M,R_N].
\end{eqnarray}
\end{subequations}
For the convenience of later analysis, we decompose the chiral gauge fields into vector and axial-vector fields,
\begin{equation}
V_M^i={1\over 2}(L_M^i+R_M^i),\qquad A_M^i={1\over 2}(L_M^i-R_M^i).
\end{equation}
Then, the strengths of the gauge fields read
\begin{subequations}
\begin{align}
F_{MN}^V=\partial_{M}V_N-\partial_{N}V_M-i[V_M,V_N]-i[A_M,A_N],\\
F_{MN}^A=\partial_{M}A_N-\partial_{N}A_M-i[V_M,A_N]-i[A_M,V_N],
\end{align}
\end{subequations}
and the covariant derivative of $X$ becomes
\begin{equation}
D_MX=\partial_{M}X-i[V_M,X]-i\{A_N,X\}.
\end{equation}

Taking the action Eq.\eqref{action} as a probe, we have the AdS-Schwarzchild black hole solution as the background geometry,
\begin{eqnarray}\label{matrics}
    ds^2=e^{2A(z)}\left (f(z)dt^2-dx^idx_i-{1\over f(z)}dz^2\right ),
\end{eqnarray}
with
\begin{eqnarray}
    A(z)=-\text{ln}(z)\qquad\text{and}\qquad f(z)=1-{z^4\over z_h^4},
\end{eqnarray}
where $z_h$ is the horizon defined by $f(z_h)=0$. The temperature of 4D system is identified as the Hawking temperature
\begin{eqnarray}\label{temperature}
    T=\left |{f'(z_h)\over {4\pi}}\right|={1\over{\pi z_h}}.
\end{eqnarray}

An IR-modified soft-wall AdS/QCD model with a modified 5D mass,
\begin{equation}\label{m5}
    m_5^2(z)=-3-\mu_c^2 z^2,
\end{equation}
was proposed in Ref.~\cite{Fang:2016nfj}, with $\mu_c$ another free parameter. Under this background, the  predicted light meson spectra are consistent with the experimental data, together with a good description of the chiral phase  transition.  This simple modification can be also considered as an effective interaction between the scalar field and the dilaton field, $\Phi\chi^2$.

According to the holographic dictionary, only the diagonal components of $X$ survive in the QCD vacuum. In this paper, we take the degenerate up ($u$) and down ($d$) quark masses and denote them as $m_q=m_u=m_d$. Then, we have
\begin{equation}\label{defchi}
    X=\frac{\chi}{2} I,
\end{equation}
with $I$ the $2\times 2$ identity matrix.
The background field $\chi$ is related to the chiral condensate. Inserting Eqs.~\eqref{matrics}-\eqref{defchi} into the action Eq.~\eqref{action}, one can derive the equation of motion (EOM) for $\chi$ as
\begin{equation}\label{eofchi}
    \chi''+\left(3A'+{f'\over f}-\Phi'\right)\chi'+{e^{2A}\over f}\left[(3+\mu_c^2 z^2)-{\lambda\chi^2\over 2}\right]\chi=0.
\end{equation}
It is a second-order nonlinear ordinary differential equation with multiple singularities, and complete analytical solutions are usually impossible. However, an analytical analysis of the chiral condensation around the critical temperature has been given in Ref.~\cite{Cao:2020ryx}. In the following, we will introduce the numerical strategy to solve it.

At the ultraviolet (UV) boundary, one can have an asymptotic expansion as
\begin{equation}\label{BC:UVchi}
    \chi(z\rightarrow 0)=m_q\zeta z+{\sigma\over \zeta} z^3+{m_q\zeta \over 4}(4\mu_g^2-2\mu_c^2+m_q^2\zeta^2\lambda)z^3 \text{ln}(z)+\mathcal{O}(z^4),
\end{equation}
where the coefficients $m_q$ and $\sigma=\langle \bar{q}q\rangle$ are interpreted as quark mass and chiral condensate, respectively,  according to the holographic dictionary. Here, $\zeta=\sqrt{N_c}/2\pi$ is a normalization constant derived by comparing the two point correlation function of $\langle \bar{q} q(k), \bar{q} q(0) \rangle$ with the result from the 4D calculation in large momentum limit~\cite{Cherman:2008eh}. The boundary expansion near the horizon can also be obtained as
\begin{eqnarray}\label{borizonofchi}
\chi(z\rightarrow z_h)= c_0 -\frac{c_0\left(2\mu _c^2z_h^2-c_0^2\lambda+6\right)}{8z_h}\left(z_h-z\right)+\mathcal{O}[(z_h-z)^2],
\end{eqnarray}
with $c_0$ an integration constant.

\begin{table}[tbp]
\centering
\begin{tabular}{|c|c|c|c|c|}
\hline
Parameter &$m_q$(GeV)&$\mu_g$(GeV)&$\mu_c$(GeV) & $\lambda$\\
\hline
Value &$3.22\times 10^{-3}$  & $0.44$ & $1.45$ & 80\\
\hline
\end{tabular}
\caption{\label{tab:parameters} The parameter values taken from~Ref.~\cite{Fang:2016nfj}. With the giving parameters, the light meson spectra,  which are consistent with the experimental data, can be solved and the chiral phase transition is also well described.}
\end{table}

\begin{figure}
  \centering
  \includegraphics[width=0.6\textwidth]{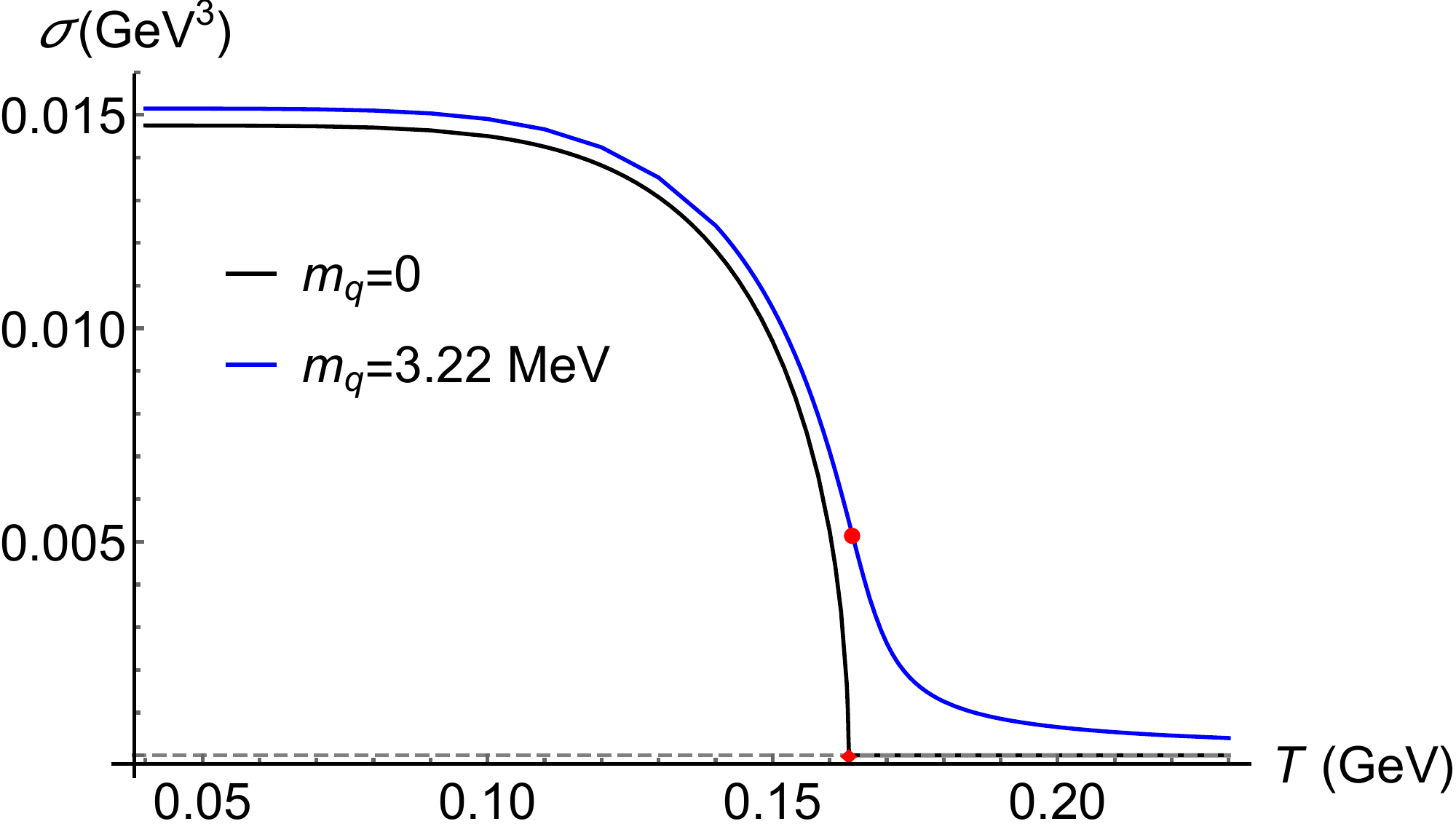}
  \caption{Chiral condensate $\sigma$ as a function of temperature $T$ in the chiral limit ($m_q=0$, the black curves) and with the physical quark mass ($m_q=3.22$, the blue curve). The red circles represent the critical point ($T_c$) of the second order phase transition and the crossover point ($T_{cp}$), which is identified as the fastest decreasing point of $\sigma$, i.e., ${d^2 \sigma}/{d T^2}|_{T=T_{cp}}=0$. Numerically,  we get $T_c=0.163$ GeV and $T_{cp}=0.164$ GeV. }\label{sigma}
\end{figure}

The model parameters, $m_q, \mu_g, \mu_c$ and $\lambda$, are listed in table~\ref{tab:parameters}, which are fitted in Ref.~\cite{Fang:2016nfj} to describe the meson spectra. It should be pointed out that the main goal of our work is to investigate the trend of the light meson masses and thermal widths at finite temperature. Thus, the analysis of this paper is mainly qualitative. Combining the asymptotic expansion solutions in Eqs.~\eqref{BC:UVchi} and \eqref{borizonofchi}, we can solve the EOM for $\chi$ in Eq.~\eqref{eofchi}, and extract the chiral condensate $\sigma$ by the ``shooting method''. The basic strategy of the algorithm is to transform the boundary problem to a iteration problem. More details can be found in our previous work~\cite{Cao:2020ryx}. In Fig.~\ref{sigma}, we show the chiral phase transition in the chiral limit and with the physical quark mass. In the chiral limit, the chiral phase transition is a second-order phase transition at a temperature $T_c=0.163$ GeV. With the physical quark mass, $m_q=3.22$ MeV, the exact chiral symmetry is broken by the small quark mass, thus the second-order phase transition becomes a crossover, and the chiral crossover temperature $T_{cp}$ equals $0.164$ GeV.

\section{Spatial correlations and screening masses}\label{III}
In this section, we will investigate the screening masses of the light mesons, including the scalar meson ($S$), the pion ($\pi$), the vector meson ($\rho$) and the axial-vector meson ($a_1$). For a 4D gauge invariant operator $\mathcal{O}$, one has the spatial correlation function as
\begin{equation}
\tilde{G}(\bold{x})\sim\langle   \hat{\mathcal{O}}(\bold{x}) \hat{\mathcal{O}}(0)\rangle.
\end{equation}
In the long-distance regime, the spatial correlation function in coordinate space would decay exponentially as
\begin{equation}\label{DF:screeningmassx}
 \tilde{G}(\bold{x})\sim e^{-m_{\rm{scr}} |\bold{x}|}/ |\bold{x}|,
\end{equation}
where $m_{\rm{scr}}$ is defined as the screening mass~\cite{DeTar:1987ar}. Generally, it is dominated by the ground state corresponding to the hadronic operator $\hat{\mathcal{O}}$. When transferred to momentum space, it becomes
\begin{equation}\label{DF:screeningmassp}
 G(\bold{p}) \sim \frac{1}{\bold{p}^2+m^2_{\rm{scr}}}.
\end{equation}
Therefore, we could extract the screening mass from the pole of the spatial correlation function.  The inverse of $m_{\rm{scr}}$ is a characteristic length scale $\zeta=m_{\rm{scr}}^{-1}$. When the distance is longer than $\zeta$, roughly, the fluctuations of $\hat{\mathcal{O}}$ are not correlated.

\subsection{Spatial correlations in the soft-wall AdS/QCD}\label{sec:GF}
It is hard to directly calculate the strong coupling correlation function in the framework of the quantum field theory. However, the holographic duality provides a possible way to handle this problem.  The basic idea of this duality is to map the strong coupling quantum field theory to the dual gravity theory, which could be easily solved. According to the holographic dictionary~\cite{Maldacena:1997re,Gubser:1998bc,Witten:1998qj,Kovtun:2004de}, the 4D operator $\mathcal{O}(x)$ and the 5D field $\phi(x,z)$ are connected through the equivalence of the partition functions,
\begin{equation}\label{equilalence}
\left.\left < e^{i\int d^4x\phi_0(x)\hat{\mathcal{O}}(x)}\right>=e^{iS_{5D}[\phi]}\right|_{\phi(x,z=0)=\phi_0(x)},
\end{equation}
where $\phi(x,z)$ is the classical solution of the equation of motion for the 5D field in the bulk, and the boundary value $\phi(x,z=0)$ equals the 4D source $\phi_0(x)$~\cite{Maldacena:1997re,Gubser:1998bc,Witten:1998qj,Kovtun:2004de}. From Eq.~\eqref{equilalence}, $\langle\hat{\mathcal{O}}(x)\hat{\mathcal{O}}(0)\rangle$ can be derived by taking the second functional derivative of $S_{5D}$ with respect to $\phi_0$. In the following, we will follow the prescription given in Ref.~\cite{Son:2002sd} to derive the spatial correlation functions of the light mesons in the IR-modified soft-wall AdS/QCD model.

\subsubsection{Scalar channel}\label{scr:scalarchannel}
The mesons are excitations above the vacuum in 4D field theory. In the dual 5D theory, they are perturbations on the background fields. For the scalar and pseudo-scalar mesons, the perturbations have the following form,
\begin{align}\label{perturbationoffields}
    X&=(\chi+S)\frac{I}{2}e^{2i\pi^it^i},
\end{align}
with $S$ representing scalar perturbation and $\pi^i$ representing pseudo-scalar perturbation.

Inserting Eq.~\eqref{perturbationoffields} into the action Eq.~\eqref{action}, one can expand the action to second-order and neglect the higher-order terms of $S$ and $\pi^i$. So the effective action of the scalar fluctuation becomes
\begin{eqnarray}\label{scalar-T}
    S_{\rm{S}}&=&\frac{1}{2}\int d x^5 \sqrt{g}e^{{-\Phi}}\bigg [ g^{\mu\nu}\partial_{\mu}S\partial_{\nu} S+g^{zz}(\partial_z S)^2-m_5^2 S^2-\frac{3\lambda}{2}\chi^2 S^2\bigg ].
\end{eqnarray}
Since we will focus on the spatial correlation function, we assume that $S$ and $\pi^i$ are static perturbations, i.e., the function of $\bold{x}$ and $z$ only.
One can transform the system from the coordinate space ($\bold{x}$) to the momentum space ($\bold{p}$) by taking the following Fourier transformation,
\begin{equation}
S(\bold{x},z)={1\over (2\pi)^3}\int  d^3\bold{p} e^{i \bold{px}}S( \bold{p},z).
\end{equation}
Without loss of generality and for simplicity, we choose $\bold{p}$ along the $x_1$-direction, i.e., $\bold{p}=(p,0,0)$. Thus, we get the EOM for the scalar meson as
\begin{equation}\label{EOM:scalar}
    S''+\left(3 A'+\frac{f'}{f}-\Phi'\right)S' -\left(\frac{p^2}{f}+\frac{2 m_5^2+3 \lambda \chi ^2}{2 f}A'^2 \right)S=0,
\end{equation}
in which the prime represents the derivative with respect to $z$.

The EOM for the scalar field, Eq.~\eqref{EOM:scalar}, is a linear second-order differential equation with multiple singularities. No analytical solution is available so far. However, we can solve it  numerically with the constraints at both the UV boundary and the horizon. We expand $S(p, z)$ around the UV boundary $z=0$, then we can derive the asymptotic expansion as
\begin{equation}\label{BC:scalarboundary}
S(z\rightarrow 0)=s_1 z+s_3 z^3-\frac{1}{4} s_1  \left[2 \left(-p^2+{\mu_c}^2-2 {\mu_g}^2\right)-3 \zeta ^2 \lambda  m_q^2\right]z^3 \log (z)+\mathcal{O}(z^4),
\end{equation}
where $s_1$ and $s_3$ are the integration constants. According to the holographic dictionary, $s_1$ corresponds to the external source $J_S$, and $s_3$ corresponds to the operator $\bar{q}q$.

 Near the horizon $z=z_h$, one can also get the non-singular expansion as
\begin{eqnarray}\label{BC:scalarhorizon}
S(z\rightarrow z_h)= s_{h0}{-\frac{6-3 c_0^2 \lambda -2 p^2 z_h^2+2 \mu_c^2 z_h^2}{8 z_h }}s_{h0}(z_h-z)+\mathcal{O}[(z_h-z)^2],
\end{eqnarray}
where $s_{h0}$ is another integration constant and $c_0$ is the integration constant of the asymptotic expansion of $\chi$ at the horizon as shown in Eq.~\eqref{borizonofchi}.  In our previous study, $c_0$ has already been solved numerically for different temperatures or $z_h$.

Substituting Eq.~\eqref{EOM:scalar} into Eq.~\eqref{scalar-T}, one can get the corresponding on-shell action as
\begin{eqnarray}\label{onshellactionofscalar}
    S_{\rm{S}}^{\rm{on}}&=&\left .-\frac{1}{2}\int  dp f(z) S(-p, z) e^{3 A(z)-\Phi (z)} S'(p, z)\right |_{z=\epsilon}^{z=z_h},
\end{eqnarray}
where $\epsilon$ is an UV cutoff regularizing the on-shell action.
Finally, one can derive the two-point spatial Green's function of  the scalar meson by taking the second-order derivative of the on-shell action $S_{\rm{S}}^{\rm{on}}$ with respect to the external source $J_S$, \footnote{The UV divergent terms, $1/z^2$ and $\log(z)$, are thrown away as a simple regularization since they are irrelevant for the mass poles and spectral functions.}
\begin{eqnarray}\label{G:scalar}
    G_{\rm{S}}(p)=\left .\frac{\delta^2 S_{\rm{S}}^{\rm{on}}}{\delta J_S^*\delta J_S}\right|_{z=\epsilon}=-\frac{4 s_3}{s_1}-\frac{3}{4}  \zeta ^2 \lambda  m^2+\frac{1}{2} \left(-2 \mu_g ^2+\mu_c^2-p^2\right).
\end{eqnarray}

From Eq.~\eqref{G:scalar}, once $s_1$ and $s_3$ are obtained, the spatial Green's function $G_S(p)$ could be calculated. One can start from the horizon boundary condition and extract $s_1, s_3$ in the UV boundary condition. Since Eq.~\eqref{EOM:scalar} is linear, the integration constant $s_{h0}$ could be set to 1 without loss of generality. Then, one can solve Eq.~\eqref{EOM:scalar} under  a given value of $p^2$ and the initial condition Eq.~\eqref{BC:scalarhorizon}. After that, $s_1$ and $s_3$ could be extracted from the UV boundary condition. They are functions of $p^2$, i.e., $s_1(p^2), s_3(p^2)$. One can obtain the screening mass from the condition $s_1(p^2)=0$, where $G_s(p)$ becomes singular. It is easy to see that Eq.~\eqref{EOM:scalar} is singular at $z=0$ and $z=z_h$. Therefore, in our numerical algorithm, we have to slightly move the UV and the horizon initial points to $z=\epsilon$ and $z=z_h-\epsilon$, respectively. But we have checked that the solutions are stable under the variation of $\epsilon$. Taking $T=0.1$ GeV and $m_q=3.22$ MeV  for an example, we get $m_{\rm{S,scr}}=1.0629760, 1.0629760, 1.0629760, 1.0632963~\rm{GeV}$ for $\epsilon=10^{-8}, 10^{-6}, 10^{-4} , 10^{-2}~\rm{GeV}^{-1}$, respectively. In the later calculations, we will take $\epsilon=10^{-8}~\rm{GeV}^{-1}$.

\subsubsection{Pseudo-scalar channel}
Being different from the scalar channel, the pion field and the longitudinal part ($\varphi^i$) of the axial-vector field are coupled in the pseudo-scalar channel. For convenience, we will take the following decomposition of the gauge field,
\begin{subequations}\label{axialgauge}
\begin{eqnarray}
    &&a_\mu^i=a_\mu^{T,i}+\partial_\mu\varphi^i,\\
    &&\partial^\mu a_\mu^{T,i}=0.
\end{eqnarray}
\end{subequations}
The pion fluctuation becomes
\begin{eqnarray}\label{actionpion}
   S_{\rm{\pi}}&=&-\frac{1}{2 {g_5}^2}\int d^5x \sqrt{g}e^{-\Phi}\sum _{i=1}^3 \bigg\{{g^{\mu \nu }} {g^{zz}} {\partial_z }{\partial_\mu \varphi^i}{\partial_z }{\partial_\nu \varphi^i} -{g_5}^2 \chi ^2 \left({g^{\mu \nu }}  \partial_\mu\varphi^i \partial_\nu\varphi^i\right.\nonumber\\
    & & \left.+{g^{\mu \nu}} \partial_\mu\pi^i \partial_\nu\pi^i+{g^{zz}} {(\partial_z\pi^i)}^2 -2 {g^{\mu \nu}}
    \partial_\mu \varphi^i \partial_\nu \pi^i\right)\bigg\}.
\end{eqnarray}

From the action Eq.~\eqref{actionpion}, one can derive the EOMs for the pion field as
\begin{subequations}\label{EOM:pion}
    \begin{eqnarray}
       &&  \varphi '' + (A'+\frac{f '}{f}- \Phi ')\varphi '-\frac{e^{2 A} g_5^2 \chi ^2}{f}\left(\varphi -\pi \right)=0,\\
      &&  \pi ^{''}+\left(3 A'+\frac{f'}{f}-\Phi '+\frac{2 \chi '}{\chi }\right)\pi^{'}+\frac{ p^2}{f}\left(\varphi-\pi \right) = 0.
\end{eqnarray}
\end{subequations}
The asymptotic solutions of the EOMs for the pion field at the UV boundary can be easily derived as
\begin{subequations}\label{BC:pionboundary}
    \begin{eqnarray}
        \varphi(z\rightarrow0)&=& c_f-\frac{1}{2} \zeta ^2 g_5^2 m_q^2 \pi_0 z^2 \log (z)+\varphi _2 z^2+\mathcal{O}(z^3),\\
        \pi(z\rightarrow 0) &=& \pi_0+c_f-\frac{1}{2} \pi_0 (\omega ^2-p^2) z^2 \log (z)+\pi _2 z^2+\mathcal{O}(z^3),
    \end{eqnarray}
\end{subequations}
where $c_f$, $\varphi_2$, $\pi_0$, and $\pi_2$ are the integration constants. As we point out in Ref.~\cite{Cao:2020ryx}, $c_f$ is a redundant free parameter and can be set to zero for simplicity. $\pi_0$ is identified as the external source $J_{\pi}$. On the other hand, we can also derive the boundary conditions at the horizon, which take the following forms,
\begin{subequations}\label{BC:pionhorizon}
    \begin{eqnarray}
    {\varphi(z\rightarrow z_h)}  &=& -\frac{c_0^2\pi^2}{z_h}\pi_{h0}(z_h-z)+\mathcal{O}[(z-z_h)^2],\\
    {\pi(z\rightarrow z_h)} &=& \pi_{h0}+\frac{ p^2 z_h}{4}\pi_{h0}(z_h-z)+\mathcal{O}[(z-z_h)^2].
    \end{eqnarray}
\end{subequations}
Here, $\pi_{h0}$  is another integration constant. The on-shell action of the pion part is
\begin{eqnarray}\label{action:pion}
    {S_{\rm{\pi}}^{\rm{on}}=  -\frac{1}{2 g_5^2}\int dp\  e^{A-\Phi }\left [e^{2 A} g_5^2 f  \chi ^2\pi(-p,z)  {\pi'}(p, z)+p^2 f\varphi(-p, z)  \varphi'(p, z) \right ] \bigg |_{z=\epsilon}^{z=z_h}}.
\end{eqnarray}
Following the prescription, we obtain the two-point spatial Green's function of the pion as
\begin{eqnarray}\label{G:pion}
    G_{\rm{\pi}}(p)= \frac{\delta^2 S_{\pi}^{\rm{on}}}{\delta J_\pi^*\delta J_\pi}=-\frac{1}{2\pi _0}\zeta ^2 m_q^2 \left[\pi _0 p^2+4 \pi _2\right]. 
\end{eqnarray}

Once one numerically gets the coefficients ($\pi_0$ and $\pi_2$) or their ratio,  the spatial correlation function would be obtained. We will leave the discussion on the numerical results in Sec.~\ref{sec:screeningmass}.

\subsubsection{Vector and axial-vector channels}
In this subsection, we turn to the spatial Green's functions of the vector and the axial-vector channels.  The process is similar to the scalar one, and we will skip some details. To get rid of the mixing of the axial-vector field and the pion field, we take the decomposition in Eq.~\eqref{axialgauge} and represent the transverse part of the axial-vector as $a_{1,\mu}=a^T_{\mu}$.  Thus, from the action Eq.~\eqref{action}, we have the vector fluctuation up to the second-order as
\begin{eqnarray}\label{action:vector}
     S_{v}=-{1\over 2g_5^2}\int d^5x \sqrt{g} e^{-\Phi}\sum_{i=1}^3\{g^{zz}g^{\mu\nu}\partial_zv_{\mu}^i\partial_zv_{\nu}^i+g^{\mu\nu}g^{mn}\partial_{\mu}v_m^i\partial_{\nu}v_n^i\},
 \end{eqnarray}
and the axial-vector fluctuation as
\begin{eqnarray}\label{action:axialvector}
    S_{a}&=&-{1\over 2g_5^2}\int d^5x \sqrt{g} e^{-\Phi}\left\{\sum_{i=1}^3\{g^{zz}g^{\mu\nu}\partial_za_{1,\mu}^i\partial_za_{1,\nu}^i+g^{\mu\nu}g^{mn}\partial_{\mu}a_{1,m}^i\partial_{\nu}a_{1,n}^i\}\nonumber\right.\\
    & &\left.-g_5^2\chi^2\sum_{i=1}^3g^{mn}a_{1,m}^ia_{1,n}^i\right\}.
\end{eqnarray}

For the spatial fluctuations, we can consider only  the spatial components  and it is convenient to take the gauge condition $v_z={a}_{1,z}=0$.
Taking the Fourier transformation, we derive the vector and axial-vector EOMs as
\begin{equation}\label{EOM:vector}
    v ''+  \left(A'+\frac{f'}{f}-\Phi '\right)v '-\frac{p^2}{f}v  =0,
\end{equation}
and
\begin{equation}\label{EOM:axialvector}
    a_{1}''+\left(A'+\frac{f'}{f}-\Phi '\right)a_{1}' -\frac{ e^{2 A} g_5^2 \chi ^2+p^2}{f}a_{1}=0,
\end{equation}
respectively.

Similarly, we obtain the UV boundary conditions for the vector,
\begin{equation}\label{BC:vectorboundary}
   v(z\rightarrow 0)=v_0+v_2 z^2+\frac{1}{2} v_0 p^2 z^2\log (z) +\mathcal{O}(z^3),
\end{equation}
and the axial-vector,
\begin{equation}\label{BC:axialvectorboundary}
    a_{1}(z\rightarrow 0)=a_{1,0}+a_{1,2} z^2+\frac{1}{2} a_{1,0}z^2 \log (z) \left[ \zeta ^2 g_5^2 m_q^2+p^2 \right]+\mathcal{O}(z^3),
\end{equation}
where $v_0, v_2, a_{1,0}$ and $a_{1,2}$ are the integration constants.

The on-shell actions of the vector and the axial-vector mesons are
\begin{equation}
    {S_{v}^{on}}= -\frac{1}{2g_5^2}\int dp e^{A(z)-\Phi (z)}v(z) f(z)  v'(z),
 \end{equation}
 and
\begin{equation}
   {S_{a}^{on}}= -\frac{1}{2g_5^2}\int dp e^{A(z)-\Phi (z)}a_1(z) f(z)  a_1'(z).
\end{equation}

Following the holographic prescription, we get the spatial Green's functions of vector and axial-vector\footnote{Here, since we have taken $\bold{p}=(p,0,0)$, the tensor structure of the current-current correlation functions are neglected. },
\begin{equation}\label{G:vector}
    {G^R_{v}(p)}=-\frac{1}{g_5^2}\left (\frac{v_2}{  v_0}+\frac{p^2}{4 }\right ),
\end{equation}
and
\begin{equation}\label{G:axialvector}
   { G^R_{a}(p)}=-\frac{1}{g_5^2}\left (\frac{a_{1,2}}{a_{1,0}}+\frac{p^2}{4}+\frac{m_q^2g_5^2\zeta^2}{4}\right).
\end{equation}

Besides, at the horizon, one has the following asymptotic expansions,
\begin{eqnarray}\label{BC:vectorhorizon}
    &&{\color{red}}v(z\rightarrow z_h)=v_{h0}+\frac{ p^2 z_h}{4}v_{h0}(z_h-z)+\mathcal{O}[(z_h-z)^2],
\end{eqnarray}
and
\begin{eqnarray}\label{BC:axialvectorhorizon}
    &&a_1(z\rightarrow z_h)=a_{h0}+\frac{ c_0^2 g_5^2+ p^2 z_h^2}{4 z_h }a_{h0}(z_h-z)+\mathcal{O}[(z_h-z)^2],
\end{eqnarray}
for the vector and the axial-vector channels, respectively. Here, $v_{h0},a_{h0}$ are the integration constants. One can solve the coefficients $v_0, v_2$ and $a_{1,0}, a_{1,2}$ numerically, and we will leave the discussion of this part later in the following section.

\subsection{Screening masses of light mesons}\label{sec:screeningmass}
In this section, we will numerically study the screening masses of the scalar meson ($m_{\rm{S,scr}}$), the pion ($m_{\rm{\pi,scr}}$), the vector meson ($m_{\rm{v,scr}}$), and the axial-vector meson ($m_{\rm{a,scr}}$). As the definition illustrated in Eq.~\eqref{DF:screeningmassx}, the screening mass can be extracted from the pole of the spatial Green's function introduced in the last subsection.

\begin{figure}[htbp]
    \centering 
        \begin{overpic}[width=0.6\textwidth]{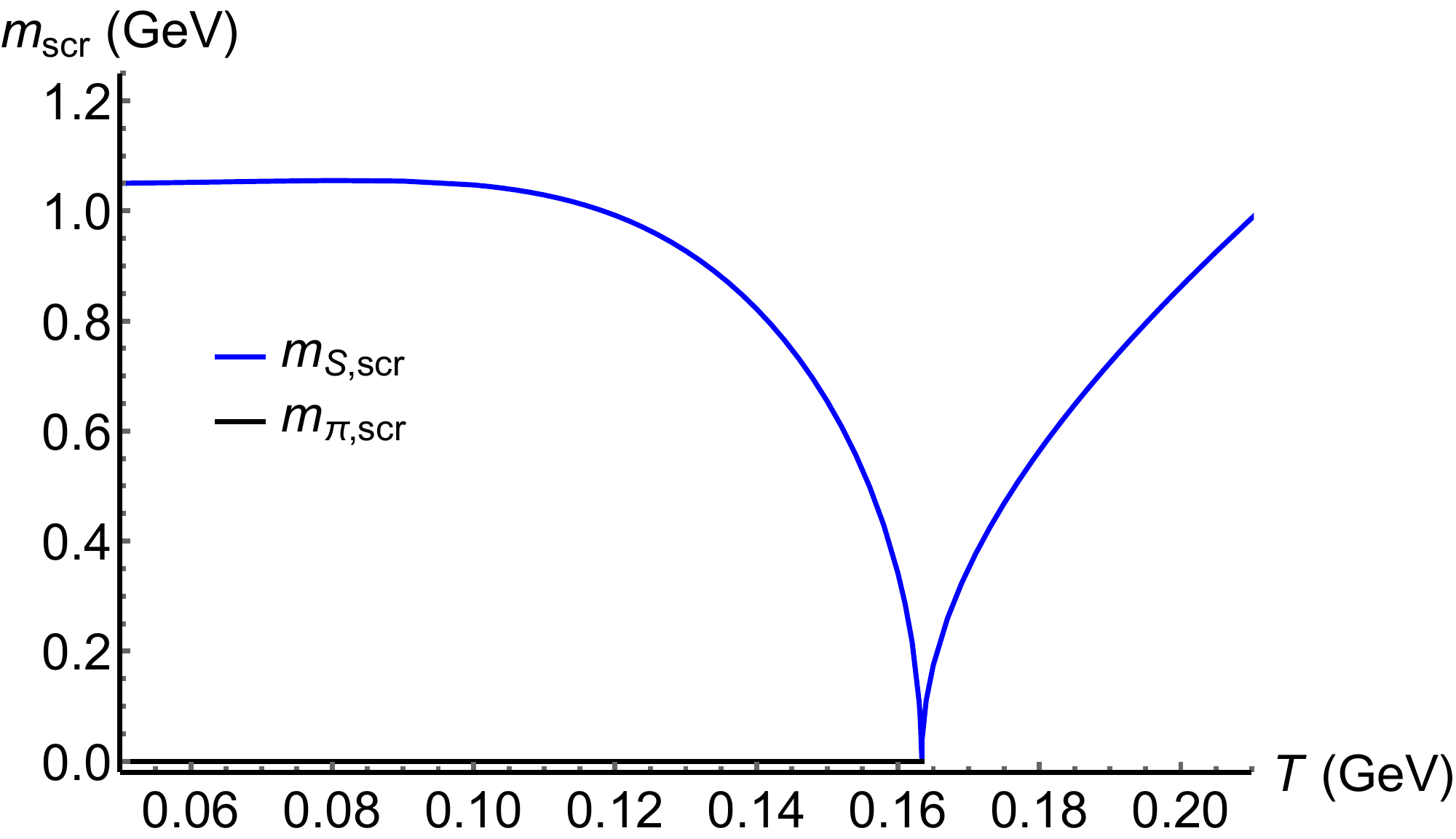}
        \end{overpic}
    \caption{\label{fig:chirallimitscalarpionscr} The temperature dependence of the screening masses of the scalar meson ($m_{\rm{S,scr}}$) and the pion ($m_{\rm{\pi,scr}}$) in the chiral limit.}
\end{figure}

As mentioned in Sec.\ref{scr:scalarchannel}, Eq.~\eqref{EOM:scalar} is a linear equation and the integration constant $s_{h0}$ in the horizon boundary condition Eq.~\eqref{BC:scalarhorizon} is an overall parameter. Thus, we could set $s_{h0}=1$.  Taking the values of the model parameters in Table.~\ref{tab:parameters}, the numerical solutions and the UV coefficients $s_1$ and $s_3$ could be easily obtained by directly integrating the linear EOM with particular values of $p^2$. One would see that for some particular values of $p^2$, $s_1$ would be zero. Actually, from the expression of the spatial correlation function, Eq.\eqref{G:scalar}, one could see that there are poles when $s_1=0$. Thus, the smallest special value of $p^2$ for getting vanishing $s_1$ is exactly the pole of the spatial correlation function, and we have $m^2_{\rm{S,scr}}= -p^2_{\rm{S},0}$. In the calculation, we find that the pole locate at a negative value of $p^2$, so  $m^2_{\rm{S,scr}}$ is always positive and well defined.

Similarly, the vector and the axial-vector meson screening masses can also be obtained, requiring $v_0=a_{1,0}=0$. However, as for the pseudo-scalar meson, the related EOMs are coupled with the longitudinal part of the axial-vector meson. Except for $\pi_{h0}=1$ and $\pi_0=0$ in Eq.~\eqref{BC:pionhorizon} and Eq.~\eqref{BC:pionboundary}, one also needs to set the redundant free parameter $c_f=0$. Then, the numerical results of the screening masses of the light mesons in the chiral limit and the physical quark mass could be calculated.

\begin{figure}[htbp]
    \centering 
    \begin{overpic}[width=0.6\textwidth]{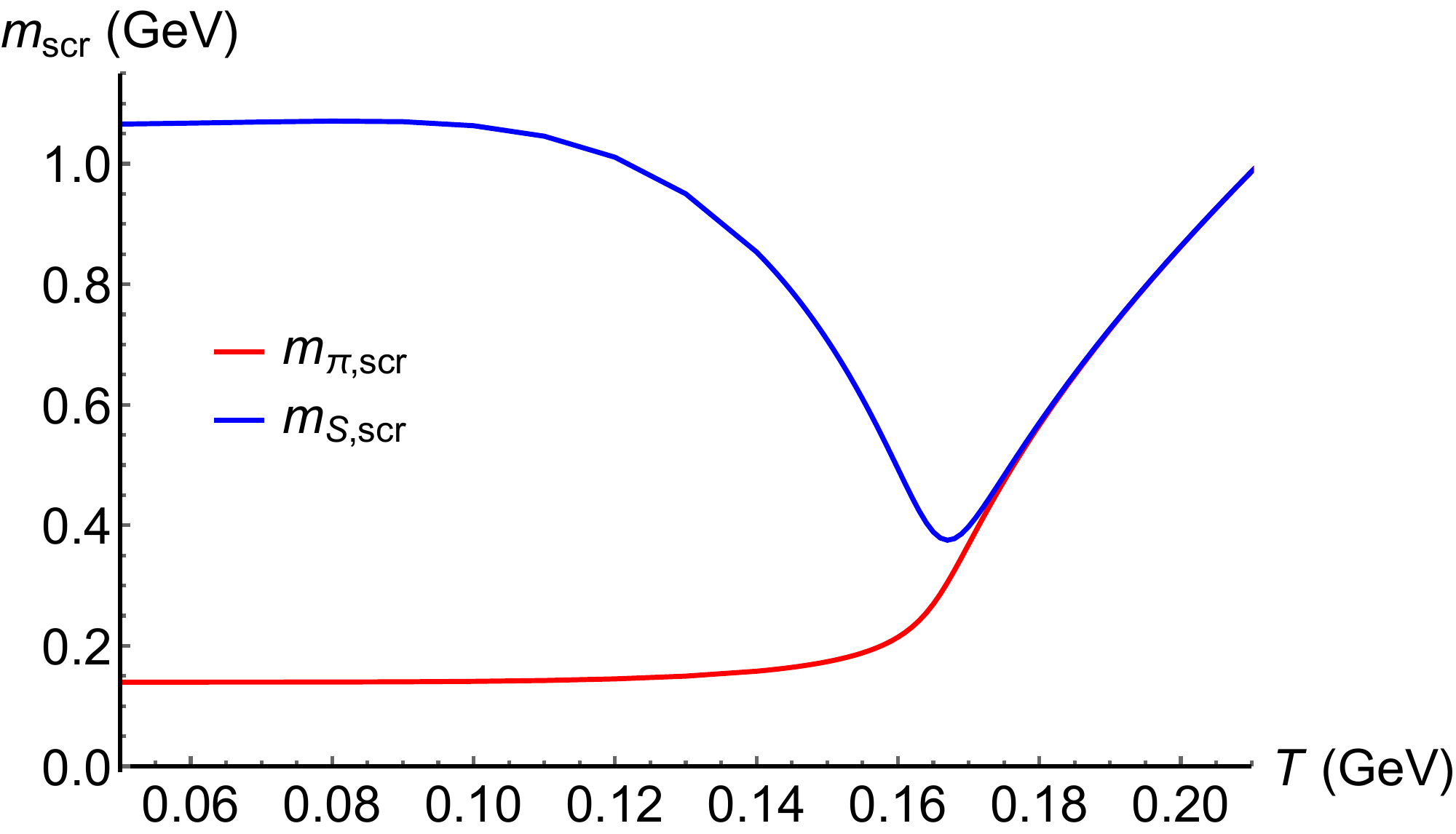}
        \put(80,50){\bf{(a)}}
    \end{overpic}
        \begin{overpic}[width=0.6\textwidth]{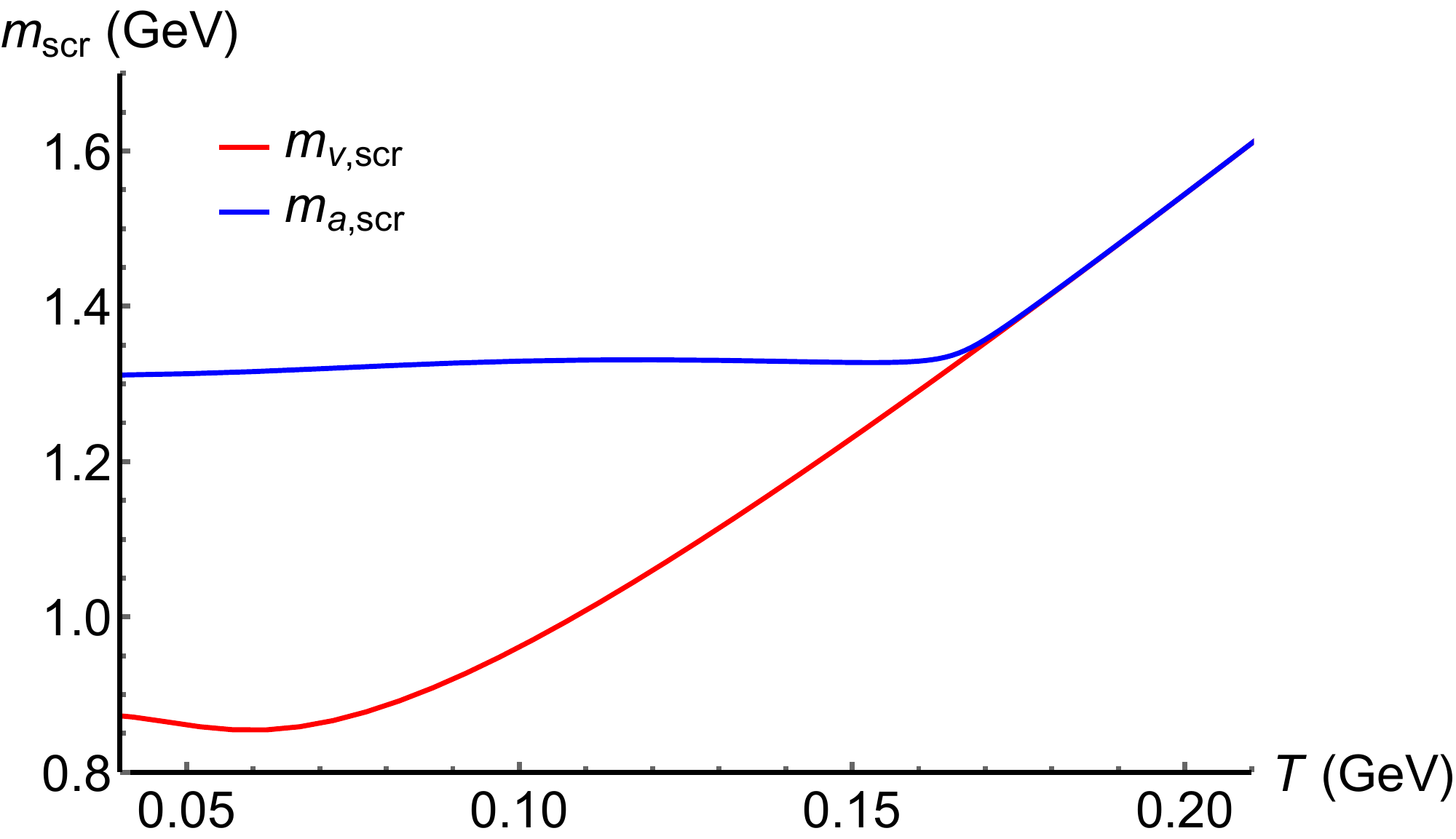}
        \put(80,50){\bf{(b)}}
    \end{overpic}
    \caption{\label{fig:screeningmass} The temperature dependence of the screening masses with the physical quark mass $m_q=3.22$MeV. (a) The screening masses of the scalar meson ($m_{\rm{S,scr}}$) and the pion ($m_{\pi,\rm{scr}}$). (b)  The screening masses of the vector meson ($m_{\rm{v,scr}}$) and the axial-vector meson ($m_{\rm{a,scr}}$),}
\end{figure}

In the chiral limit, we take $m_q=0$ to check the theoretical consistency of the model. The numerical results of the screening masses of the scalar and the pseudo-scalar mesons are shown in Fig.~\ref{fig:chirallimitscalarpionscr}. From the figure, it could be seen that the pion screening mass is precisely zero below $T=0.163 ~\rm{GeV}$. It shows the infinite correlation length $\zeta_{{\pi}}=m_{\rm{\pi,scr}}^{-1}=\infty$ of the Goldstone mode. However, the screening mass of the scalar meson decreases from $1.05 ~\rm{GeV}$ to zero with the increasing temperature for $T<0.163 ~\rm{GeV}$. These phenomena agree with the chiral restoration extracted from the chiral condensate, and the critical temperature is confirmed as  $T_c=0.163$ GeV. The chiral symmetry is spontaneously broken below $T_c$, and the pion is exactly the Nambu-Goldstone boson for the chiral phase transition. Since no chiral condensate formed above $T_c$, the EOMs for the pion in Eqs.~\eqref{EOM:pion}(a) and (b) are not applicable anymore. The form of the fluctuation for the pion in Eq.~\eqref{perturbationoffields} is not available.  It should be changed to the same form as the scalar meson~\cite{Cao:2020ryx}. It means that the pion and the scalar meson degenerate, and they share the same EOM. In this region, the screening mass monotonically increases with the increasing temperature. We proved that the screening mass increases linearly with $T$ at extremely high temperature in Appendix \ref{high temperature limit}.  Besides, we numerically calculate the critical exponent of the correlation length and get $\nu\approx 0.5$ (with the definition of $\nu$ as $\zeta_{\pi}=(T-T_c)^{-\nu}$), which agrees with the mean-field result of the 3D Ising model. To go beyond the mean-field approximation, we suggest that the back-reaction of the higher power of scalar potential and the temperature dependence of the dilaton field should be considered in a full back-reaction model~\cite{Chen:2018msc}.

Now, we consider the case with the physical quark mass, $m_q=3.22$ MeV. Since the small non-zero quark mass gives a minor explicit symmetry breaking, the second-order phase transition becomes a crossover. The pion obtains small mass $m_{\pi}(T=0)\approx 0.139$ GeV and acts as a pseudo-Goldstone boson. We numerically calculate the screening mass. The screening masses of the pion and the scalar meson are shown in Fig.~\ref{fig:screeningmass}(a). The screening masses equal the meson masses at zero temperature $m_{\pi,\rm{scr}}(T=0)= m_{\pi}(T=0)$ and $m_{\rm{S, scr}}(T=0)= m_{\rm{S}}(T=0)\approx 1.06$ GeV. With the increasing of the temperature, the screening mass of the scalar meson decreases and reaches its minimal value at $T_{S,2}=0.167 \rm{GeV}$, which is very close to the crossover temperature $T_{cp}\approx0.164$ GeV. It means that even with the physical quark mass, the screening mass would be a signal for the chiral crossover. The pion screening mass increases slowly  below $T_{cp}$, then increases sharply  and merges with the scalar one above $T_{cp}$. We note that the above results are in good agreement with the LQCD~\cite{Cheng:2010fe} and the NJL results~\cite{Jiang:2011aw}. These results confirm that the temperature-dependent behavior of screening masses of the pion and the scalar meson are strongly coupled with the chiral phase transition.

The screening masses of the vector and axial-vector mesons are given in Fig.~\ref{fig:screeningmass}(b).  It could be seen that  the screening masses of vector meson and axial-vector meson at very low temperature are very close to their in-vacuum values. Below $T_{cp}$, the curve for screening mass of the vector meson behaves a concave shape with a minimum at $T\approx0.062$ GeV, then increases monotonically.  The screening mass of the axial-vector meson has a very slight bump (almost increases monotonically). It might be caused by the dissociation effect and the vanishing of the chiral condensation at a relatively high temperature. The dissociation effect increases the screening mass, while the vanishing of chiral condensate decreases the screening mass. Above $T_{cp}$, it is interesting to see that the screening masses of the vector and the axial-vector mesons also merge with each other in the chiral symmetric phase. Also, all these behaviors of the screening mass are qualitatively consistent with the LQCD simulations in Ref.~\cite{Cheng:2010fe}.

\section{Pole masses,  thermal widths, temporal correlations,  and quasi-normal modes}\label{IV}
The pole masses of the scalar meson and pion in the current model have been extracted from the spectral functions in our previous work~\cite{Cao:2020ryx}. It has been shown that the pole masses of the scalar and pseudo-scalar mesons degenerate at temperature above the chiral crossover temperature $T_{cp}$. This could be considered as a reflection of the chiral symmetry restoration at the hadronic spectrum level. Also, it could be seen that the widths of the resonance peaks increase rapidly with the increasing temperature. It becomes hard to extract the masses and widths at high temperatures since the peaks are too broad to be considered as resonance peaks. In this situation, as shown in Ref.~\cite{Miranda:2009uw}, one can map the masses and widths to the complex frequencies, $\omega=\omega_{\rm{Re}}-i\omega_{\rm{Im}}$, of the corresponding quasi-normal modes (QNMs). In this scenario, one can also study the thermal widths of the light mesons, which are the essential quantities to describe meson dissociations.  In the holographic duality, it has been established that the QNM frequency is related to the process of thermalization in the 4D strong coupling field theory~\cite{Horowitz:1999jd,KalyanaRama:1999zj}. The real part ($\omega_{\rm{Re}}$) is the oscillation frequency of the mode, and the imaginary part ($\omega_{\rm{Im}}$) contributes to the damping rate. The real part of the lowest QNM frequency is related to the pole mass, and the corresponding imaginary part is related to the thermal width. It is also interesting to extend the study of the scalar and pseudo-scalar mesons to the vector and axial-vector sectors and get more knowledge of the thermal properties of the light mesons. We will give a detailed analysis on this topic in this section.

\subsection{Temporal correlations and pole masses of light mesons}
From the definition, the pole masses of the scalar, the pion, the vector and the axial-vector mesons  are the real parts of  the frequencies of the temporal correlation (Green's) function poles with zero spatial momentum. The main difference to extract the temporal correlation functions is that we need to consider the temporal perturbations of the corresponding fields. In this case, they depend on the frequency ($\omega$) other than the spatial momentum ($\bold{p}$), i.e., $S(\omega,z)$, $\pi(\omega, z), \varphi(\omega,z)$, $v(\omega,z)$, and $a_1(\omega,z)$. It is easy to repeat the previous derivations and get the temporal EOMs. The EOM for the scalar meson takes the form
\begin{equation}\label{EOM:scalar-omega}
    S''+\left(3 A'+\frac{f'}{f}-\Phi'\right)S' +\left(\frac{\omega ^2}{f^2}-\frac{2 m_5^2+3 \lambda    \chi ^2}{2 f}A'^2 \right)S=0.
\end{equation}
The pseudo-scalar meson would obey the following coupled equations
\begin{subequations}\label{EOM:pion-omega}
    \begin{eqnarray}
       &&  \varphi^{''} + (A'- \Phi ')\varphi ^{'}-\frac{e^{2 A} g_5^2 \chi ^2}{f}\left(\varphi -\pi \right)=0,\\
      &&  \pi ^{''}+\left(3 A'+\frac{f'}{f}-\Phi '+\frac{2 \chi '}{\chi }\right)\pi^{'}-\frac{\omega ^2 }{f^2}\left(\varphi-\pi \right) = 0.
\end{eqnarray}
\end{subequations}
The EOM for the vector meson has a simpler form
\begin{equation}\label{EOM:vector-omega}
    v ''+  \left(A'+\frac{f'}{f}-\Phi '\right)v '+\frac{\omega ^2}{f^2}v  =0,
\end{equation}
while the axial-vector meson takes a slightly different form
\begin{equation}\label{EOM:axialvector-omega}
    a_{1}''+\left(A'+\frac{f'}{f}-\Phi '\right)a_{1}' -\left[\frac{ e^{2 A} g_5^2 \chi ^2}{f}-\frac{ \omega ^2}{f^2}\right]a_{1}=0.
\end{equation}
From the above EOMs, one could derive the corresponding asymptotic expansions at the UV boundary. The expansion of the scalar sector becomes
\begin{equation}\label{BC:scalarboundary-omega}
S(z\rightarrow 0)=s_1 z+s_3 z^3-\frac{1}{4} s_1  \left[2 \left(\omega ^2+{\mu_c}^2-2 {\mu_g}^2\right)-3 \zeta ^2 \lambda  m_q^2\right]z^3 \log (z)+\mathcal{O}(z^4).
\end{equation}
That of the pseudo-scalar sector reads
\begin{subequations}\label{BC:pionboundary-omega}
    \begin{eqnarray}
        \varphi(z\rightarrow0)&=& c_f-\frac{1}{2} \zeta ^2 g_5^2 m_q^2 \pi_0 z^2 \log (z)+\varphi _2 z^2+\mathcal{O}(z^3),\\
        \pi(z\rightarrow 0) &=& \pi_0+c_f-\frac{1}{2} \pi_0 \omega ^2 z^2 \log (z)+\pi _2 z^2+\mathcal{O}(z^3).
    \end{eqnarray}
\end{subequations}
The vector sector becomes
\begin{equation}\label{BC:vectorboundary-omega}
   v(z\rightarrow 0)=v_0+v_2 z^2-\frac{1}{2} v_0 \omega ^2 z^2\log (z) +\mathcal{O}(z^3),
\end{equation}
and the axial-vector sector becomes
\begin{equation}\label{BC:axialvectorboundary-omega}
    a_{1}(z\rightarrow 0)=a_{1,0}+a_{1,2} z^2+\frac{1}{2} a_{1,0}z^2 \log (z) \left[ \zeta ^2 g_5^2 m_q^2- \omega ^2\right]+\mathcal{O}(z^3).
\end{equation}
Then we have the temporal correlation functions of the scalar, the pseudo-scalar, the vector and the axial-vector mesons as\footnote{Since only the time-dependence is considered, the tensor structure of vector and axial-vector mesons are also omitted. }
\begin{eqnarray}\label{G:scalar-omega}
    G^{\rm{R}}_{\rm{S}}(\omega)=-\frac{4 s_3}{s_1}-\frac{3}{4}  \zeta ^2 \lambda  m^2+\frac{1}{2} \left(-2 \mu_g ^2+\mu_c^2+\omega ^2\right),
\end{eqnarray}

\begin{eqnarray}\label{G:pion-omega}
    G^{\rm{R}}_{\rm{\pi}}(\omega)=\frac{1}{2\pi _0}\zeta ^2 m_q^2 \left[\pi _0 \omega ^2-4 \pi _2\right], 
\end{eqnarray}

\begin{equation}\label{G:vector-omega}
    G^R_{v}(\omega)=-\frac{1}{g_5^2}\left (\frac{v_2}{  v_0}-\frac{\omega ^2}{4 }\right ),
\end{equation}

\begin{equation}\label{G:axialvector-omega}
    G^R_{a}(\omega)=-\frac{1}{g_5^2}\left (\frac{a_{1,2}}{a_{1,0}}-\frac{\omega ^2}{4}+\frac{m_q^2g_5^2\zeta^2}{4}\right).
\end{equation}

By solving the EOMs and extracting the UV coefficients, one could get the temporal Green's functions. However, before that, one has to specify the boundary conditions at the horizon, and they would be different from those in the spatial cases. These differences mainly come from the $f^2$ factor in the denominators of the EOMs. In the spatial cases, this factor is $f$ other than $f^2$. Such a difference would cause variations of the asymptotic expansions at the horizon. The regular power expansions should be replaced with the incoming or outgoing boundary conditions. According to the prescription proposed in Ref.~\cite{Son:2003et}, the incoming conditions are for retarded Green's functions while the outgoing conditions are for advanced Green's functions. In this work, we will consider the retarded Green's functions and take the incoming conditions. The asymptotic expansions at the horizon for these four sectors would become
\begin{eqnarray}\label{BC:scalarhorizon-omega}
S(z\rightarrow z_h)= (z_h-z)^{-i\frac{\omega}{4\pi T}}\left\{s_{h0}+(z_h-z)s_{h1}+\mathcal{O}[(z_h-z)^2]\right\},
\end{eqnarray}
\begin{subequations}\label{BC:pionhorizon-omega}
    \begin{eqnarray}
    \varphi(z\rightarrow z_h)  &=& \left(z_h-z\right)^{-\frac{i \omega}{4\pi T}}\bigg\{ \frac{16 \pi ^2 c_0^2 \pi_{h0} }{\omega  z_h^2 (z_h\omega +4 i)}(z_h-z)+\mathcal{O}[(z-z_h)^2]\bigg\}+c_{h0},\\
    \pi(z\rightarrow z_h) &=& \left(z_h-z\right)^{-\frac{i \omega}{4\pi T}}\bigg\{\pi_{h0}+\pi_{h1}(z_h-z)+\mathcal{O}[(z-z_h)^2]\bigg\}+c_{h0},
    \end{eqnarray}
\end{subequations}
\begin{eqnarray}\label{BC:vectorhorizon-omega}
    &&v(z\rightarrow z_h)=(z-z_h)^{-{i\omega\over 4\pi T}}\left\{v_{h0}+v_{h1}(z_h-z)+\mathcal{O}[(z_h-z)^2]\right\},
\end{eqnarray}
and
\begin{eqnarray}\label{BC:axialvectorhorizon-omega}
    &&a_1(z\rightarrow z_h)=(z-z_h)^{-{i\omega\over 4\pi T}}\left\{a_{h0}+a_{h1}(z_h-z)+\mathcal{O}[(z_h-z)^2]\right\},
\end{eqnarray}
with
$$s_{h1}=\frac{-12+6c_0\lambda+6i z_h\omega +8 i z_h^3 \mu_g \omega-z_h^2(4\mu_c^2+3\omega^2)}{8z_h(2-z_h\omega i)} s_{h0},$$
$$\pi_{h1}=\frac{\left [z_h(-8 z_h\mu_g^2+2z_h\mu_c^2-3 \omega i)+c_0^2\left (-\lambda+\frac{16 i\pi^2}{4 i +z_h\omega}\right)\right ]\omega}{8(2 i+z_h\omega)}\pi_{h0},$$
$$v_{h1}=\frac{(2-8 z_h^2\mu_g^2-3 i z_h \omega)\omega}{8(2 i +z_h \omega)}v_{h0},$$
and
$$a_{h1}=\frac{4 i c_0^2 g_5^2+(2-8 z_h^2\mu_g^2-3 i z_h \omega)z_h\omega}{8 z_h(2 i+z_h\omega)}a_{h0}.$$

The poles of the Retarded Green's functions are located at specific frequency values, which satisfy the vanishing conditions of $s_1, \pi_0, v_0$ and  $a_{1,0}$. There are slight differences between the spatial cases and the temporal cases. From our calculation, at finite temperatures no real frequency values could satisfy these conditions. The poles are in the complex frequency plane with $\omega=\omega_{\rm{Re}}-i\omega_{\rm{Im}}$. The real parts ($\omega_{\rm{Re}}$) of the poles characterize the oscillation rates while the imaginary parts ($\omega_{\rm{Im}}$) describing the damping rates. The corresponding modes are called QNMs in the literature. Thus the complex QNM frequencies are the poles of the temporal retarded Green's functions~\cite{Miranda:2009uw}.

From the definition, the pole masses are the real parts of the temporal poles, i.e., $m_{\rm{pole}}=\omega_{\rm{Re}}$,  and the thermal widths are related to the imaginary parts by $\Gamma/2=\omega_{\rm{Im}}$. By solving the EOMs with the above boundary conditions, we can obtain the pole masses as well as the thermal widths from the retarded Green's functions of the scalar meson Eq.~\eqref{G:scalar-omega}, the pion Eq.~\eqref{G:pion-omega}, the vector meson Eq.~\eqref{G:vector-omega} and the axial-vector meson Eq.~\eqref{G:axialvector-omega}. We can take a similar numerical algorithm, and get the pole masses from the QNM frequencies.  For example, the only difference for the scalar sector is to replace Eqs.~\eqref{EOM:scalar} and \eqref{BC:scalarhorizon} with Eqs.~\eqref{EOM:scalar-omega} and \eqref{BC:scalarhorizon-omega}. For more details about the numerical algorithm, please refer to Sec.~\ref{scr:scalarchannel}.

\begin{figure}[htbp]
    \centering 
        \begin{overpic}[width=0.6\textwidth]{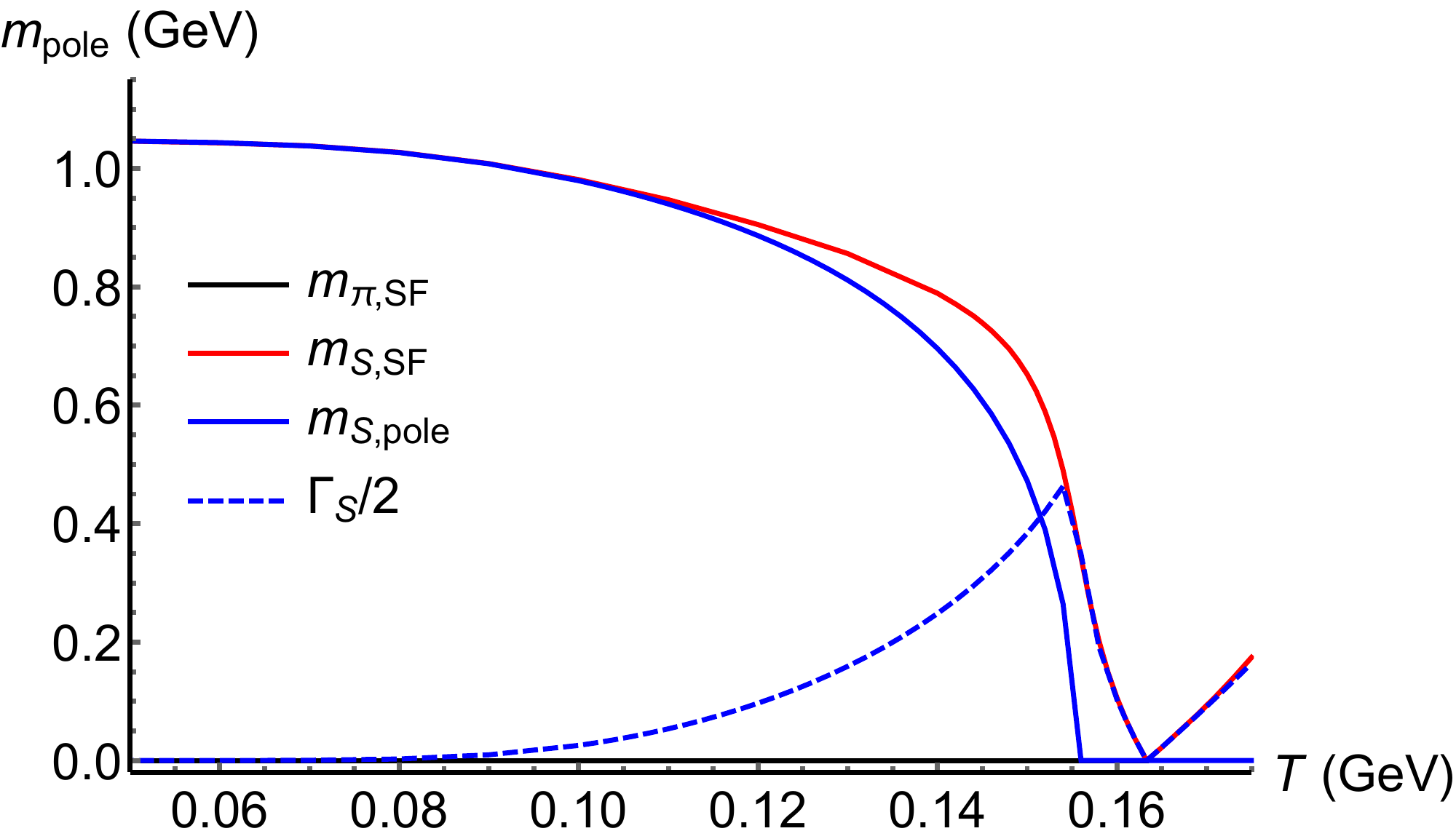}
        \end{overpic}
    \caption{\label{fig:chirallimitscalarpion} The temperature dependence of the pole masses $m_{S,pole}$  and the thermal widths $\Gamma_S$  of the scalar meson in the chiral limit. Taken from Ref.~\cite{Cao:2020ryx}, the quasiparticle masses of the scalar meson $m_{S,SF}$ and the pion $m_{\pi,SF}$, are extracted from the peak locations of the spectral functions.}
\end{figure}

Firstly, we also check the theoretical consistency by considering the chiral limit. We obtain the pole masses and the thermal widths of the scalar meson and the pion, as shown in Fig.~\ref{fig:chirallimitscalarpion}.
In the figure, the blue line and the blue dashed line represent the pole mass ($m_{\text{S,pole}}$) and the width ($\Gamma_{S}$) of the scalar meson, which are extracted from the QNM frequency. The red and black solid lines represent the pole masses of the scalar meson ($m_{\rm{S,SF}}$) and the pion ($m_{\rm{\pi,SF}}$), respectively. They are extracted from the locations of the peaks in the spectral functions and are taken from our previous study~\cite{Cao:2020ryx}. We will call them quasiparticle masses later. The critical temperature of the chiral phase transition $T_{c}$ is equal to $0.163$ GeV, at which both the pole mass and the width equal zero. Below $T_c$, the pole mass and the width of the pion are zero, which are the same as the spectral function results. However, the pole mass of the scalar meson from the QNM frequency decreases to zero at $T_{\rm{S,0}}=0.156$GeV. The width increases with the temperature and reach their maximum at $T_{\rm{S,0}}$. Then it decrease to zero at $T_c$. The pole masses from QNM frequencies are consistent with the results from spectral functions in the low-temperature region. At higher temperatures, the values of the widths gradually drive the quasiparticle masses away from the pole masses from QNM frequencies. Numerically, the relation between the quasiparticle masses and the pole masses from QNM frequencies approximately satisfies  ${m}_{S,SF}^2\approx m_{\rm{S,pole}}^2+(\Gamma_{\rm{S}}/2)^2$. We have noticed that a similar relation for the effective masses of $Z^0$ is analytically studied in Ref.~\cite{Sirlin:1991fd}. It is not difficult to comprehend this behavior.  At low temperatures, the widths are relatively negligible, ($\Gamma/2<<m_{\rm{pole}}$), and the poles are very close to the real $\omega$-axis. Therefore, the values of the locations of the spectral function peaks are almost equal to the pole masses from QNM frequencies. However, when the temperature is high, the large thermal widths drive the poles away from the real $\omega$ axis and cause significant differences between the masses from spectral functions and QNM frequencies. When the temperature is above $T_{S,0}$, since the real parts of the QNM frequencies are zero, the quasiparticle masses are dominated by the imaginary parts. Understanding the underlying reasons for the transition will be left for the future.

\begin{figure}[thbp]
    \centering 
        \begin{overpic}[width=0.45\textwidth]{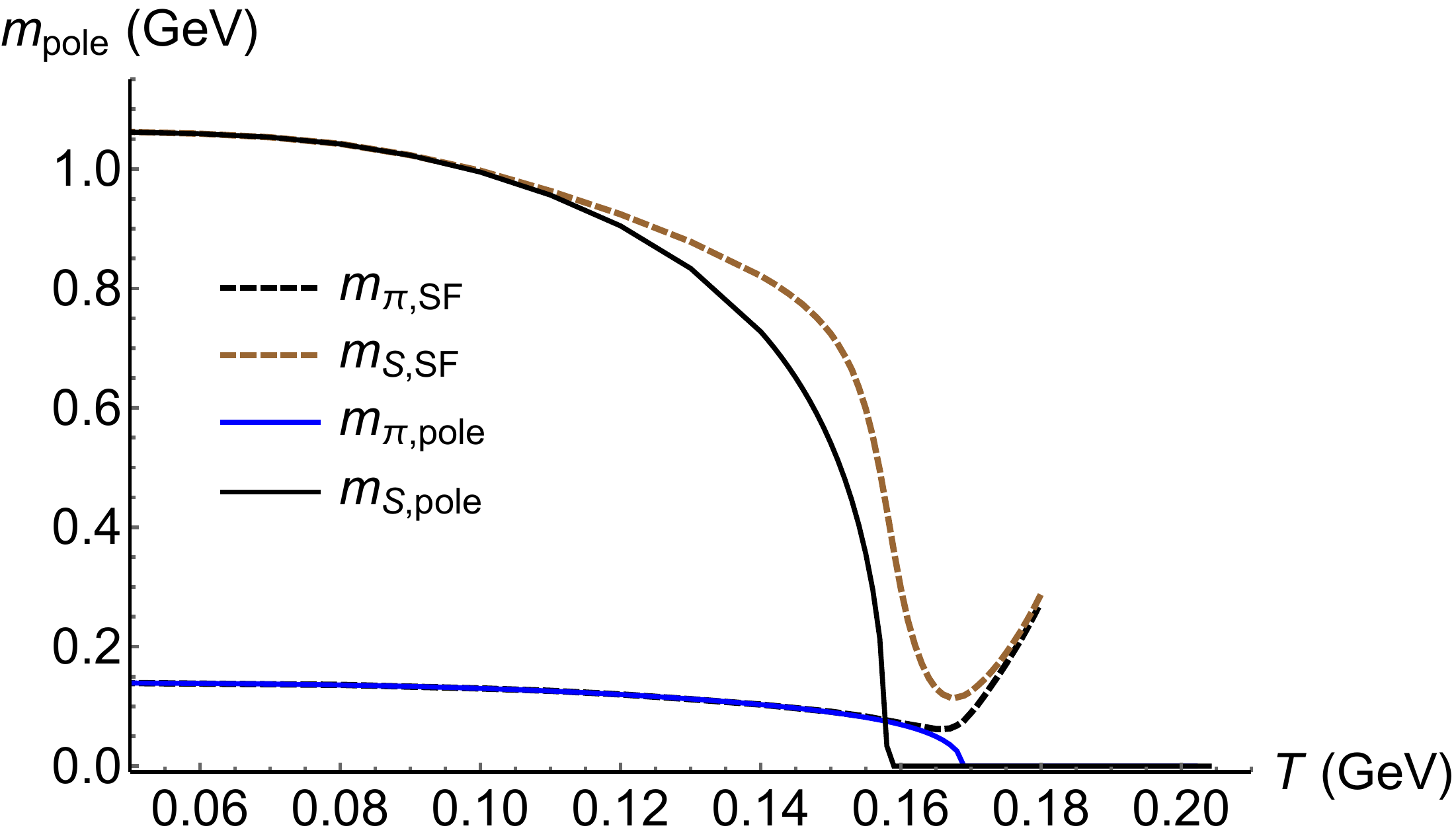}
        \put(80,50){\bf{(a)}}
    \end{overpic}
            \begin{overpic}[width=0.45\textwidth]{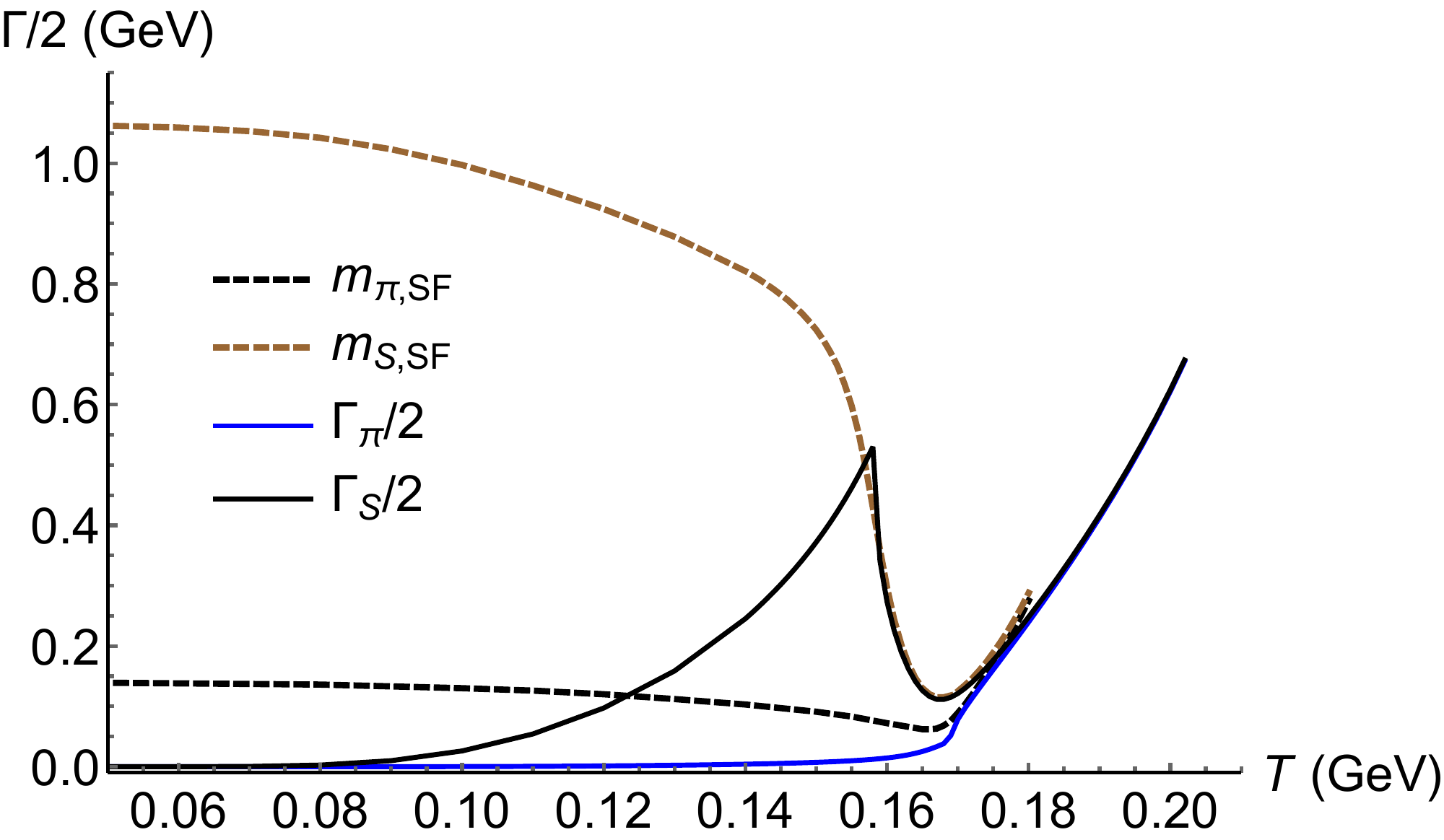}
        \put(80,50){\bf{(b)}}
    \end{overpic}
            \begin{overpic}[width=0.45\textwidth]{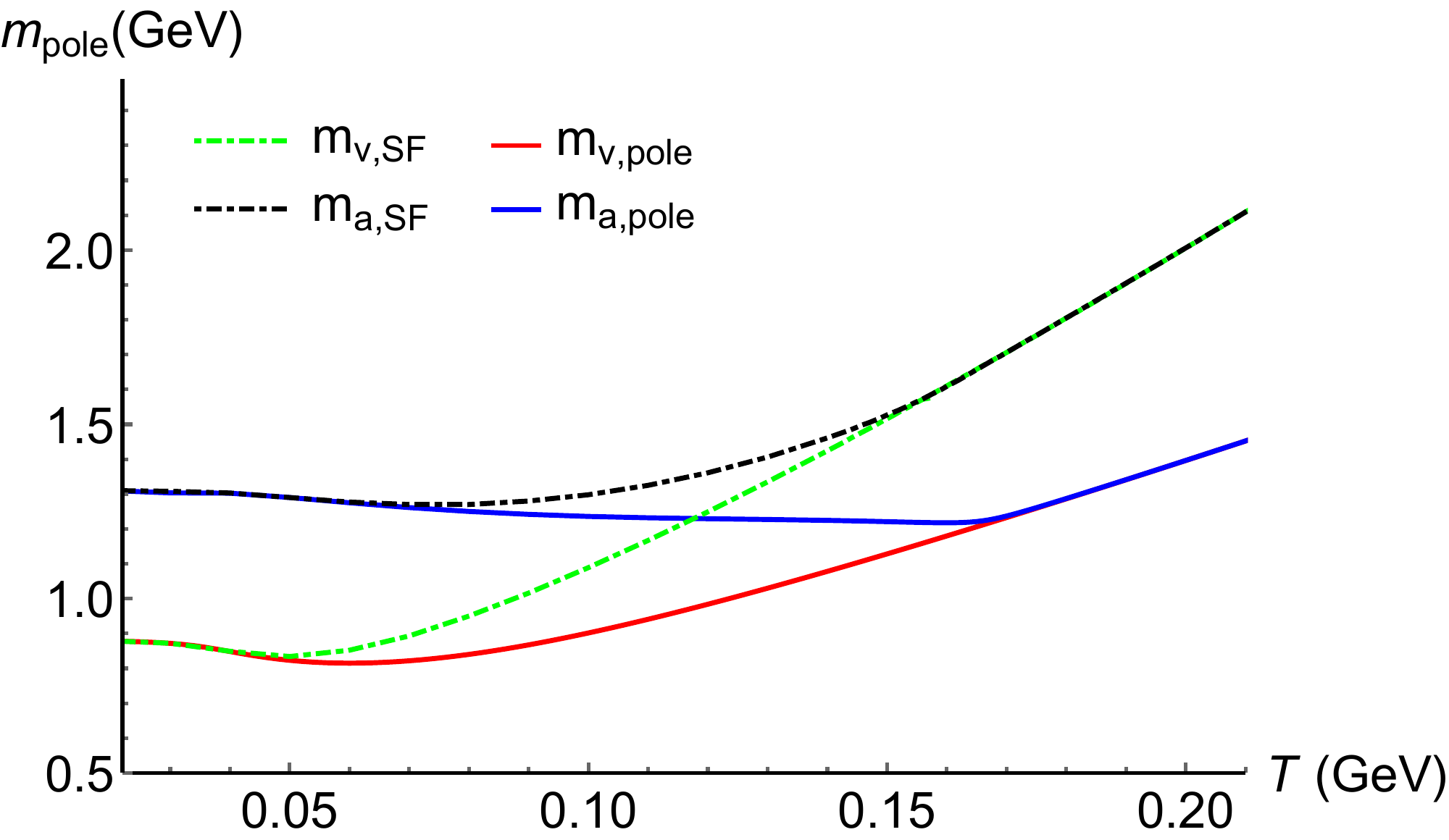}
        \put(80,50){\bf{(c)}}
    \end{overpic}
    \begin{overpic}[width=0.45\textwidth]{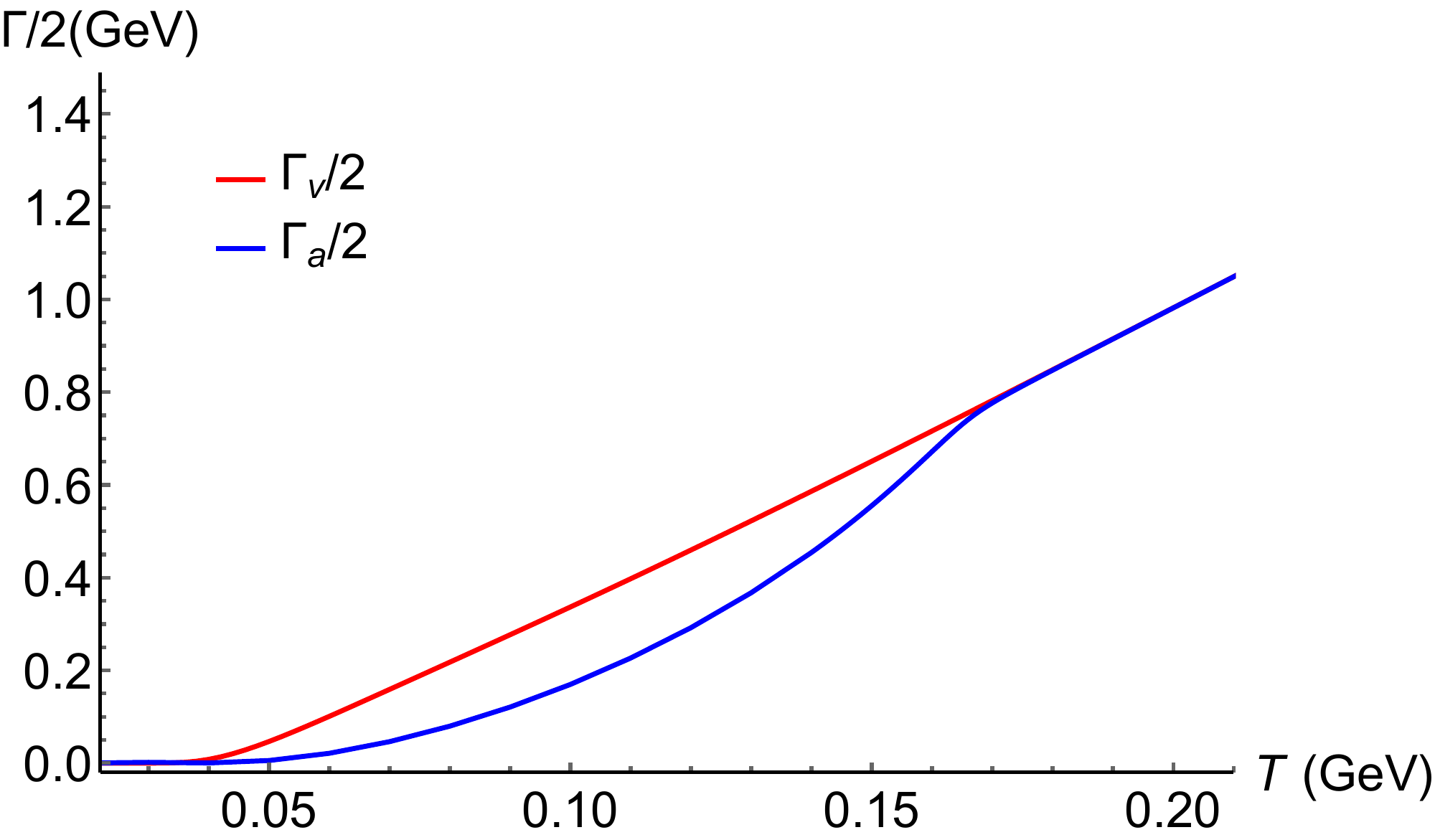}
        \put(80,50){\bf{(d)}}
    \end{overpic}
    \caption{\label{fig:polemass} The dependence of the pole masses and the widths on temperature with physical quark mass $m_q=3.22$ MeV. Pole masses (a) and thermal widths (b) of the scalar meson ($m_{\rm{S,pole}}$, $\Gamma_{\rm{S}}$) and pion  ($m_{\rm{\pi,pole}}$, $\Gamma_{\rm{\pi}}$). Pole masses (c) and thermal widths (d) of the vector meson  ($m_{\rm{v,pole}}$, $\Gamma_{\rm{v}}$) and axial-vector meson  ($m_{\rm{a,pole}}$, $\Gamma_{\rm{a}}$). The quasiparticle masses of the scalar meson and the pion extracted from the spectral functions, taken from Ref.~\cite{Cao:2020ryx}, are shown as the dashed lines in (a) and (b).}
\end{figure}

For the physical quark mass case, $m_q=3.22$ MeV, we show the pole masses and the thermal widths in Fig.~\ref{fig:polemass}. In Fig.~\ref{fig:polemass}(a), the pole mass from the QNM frequency of the pion, $m_{\rm{\pi, pole}}$, monotonically decreases with the increasing temperature and reach zero at $T_{\rm{\pi},1}=0.169$ GeV. The pole mass of the scalar meson from the QNM frequency  also monotonically decreases with the increasing temperature and reach zero at $T_{\rm{S},1}=0.158$ GeV. These are pretty similar to the results in the chiral limit. Also, the quasiparticle masses from spectral functions are close to the pole masses from the QNM frequencies at low temperature.

In Fig.~\ref{fig:polemass}(b), the comparison of thermal widths with the quasiparticle masses from spectral functions is given. It is shown that the thermal width of the pion ($\Gamma_{\rm{\pi}}/2$) increases monotonically  with the increasing temperature. However, the width of the scalar meson increases quickly below $T<T_{\rm{S},1}$, then turn to decrease, and finally approach the pion width above $T_{\rm{\pi},1}$. It could be seen that, at high temperature, the quasiparticle masses are very close to the imaginary parts of the QNM frequencies.  It indicates that the quasiparticle masses describe the pion and the scalar meson better when the imaginary parts dominate the QNM frequencies. At high temperature, the masses of the pion and the scalar meson degenerate, revealing the restoration of chiral symmetry at the hadronic spectrum level. The rapid increase of the thermal widths might be related to the dissociations of mesons.

The pole masses ($m_{\rm{v/a,pole}}$) from the QNM frequencies and quasiparticle masses ($m_{\rm{v/a,SF}}$) from the spectral functions of the vector meson and the axial-vector meson are compared in Fig.~\ref{fig:polemass}(c). It could be seen that at low temperature $m_{\rm{pole}}$ and $m_{\rm{SF}}$ are very close, while apparent deviations appear at high temperature. We will compare these two scenarios in the next section. However, the qualitative behaviors of them are very similar. The two sectors possess different values of pole masses in the chiral asymmetric phase. When the chiral symmetry is restored at a sufficiently high temperature, $T>T_{cp}$, they will merge for the degeneration of the vector and axial-vector mesons. The pole mass of the vector meson, $m_{\rm{v, pole}}$, has a slight drop from $0.88$ GeV at $T=0$ to its minimum $0.814$ GeV at $T_{v,m}\approx0.06$ GeV. The mass shift is less than $7.5\%$, which is well consistent with the predictions from the experiment and the theoretical approach in Ref.~\cite{Arnaldi:2006jq,Rapp:2009yu}. The pole mass of the axial-vector meson, $m_{\rm{a_1, pole}}$, also decreases from $1.31$ GeV at $T=0$ to its minimum $1.22$ GeV at $T_{a,c}\approx 0.16$ GeV $\approx T_{cp}$. In Fig.~\ref{fig:polemass}(d), we plot the thermal widths of the vector and axial-vector mesons. Below $T_{a,c}$,  the thermal widths of the vector and axial-vector mesons increase with the increasing temperature, while the width of the axial-vector meson is always smaller than that of the vector meson. The large values of thermal widths above $T_{cp}$ indicate the dissociations of the mesons around this temperature. However, the exact dissociation temperature requires further study, and we leave it for future. Above $T_{a,c}$, the vector and axial-vector meson masses degenerate, and their pole masses from the QNM frequencies together with the thermal widths are all monotonically increasing. The degeneration of the two sectors at high temperature is a signal of the chiral symmetry restoration.

\subsection{Spectral functions and QNMs}
In the last subsection, we have studied the pole masses from the QNM frequencies of the light mesons and have given a comparison with the results from the spectral functions. We find that these two different approaches give almost the same values at low temperature. However, the effects of the thermal width can not be ignored at high temperature. The underlying connection between these two different methods should be interesting. We will also study the spectral functions of the vector and axial-vector mesons. What is more, we will numerically verify the relationship between the pole masses from the QNM frequencies and from the spectral functions.

\subsubsection{Spectral functions for the vector and axial-vector mesons}

  \begin{figure}[thbp]
    \centering 
    \begin{overpic}[width=0.45\textwidth]{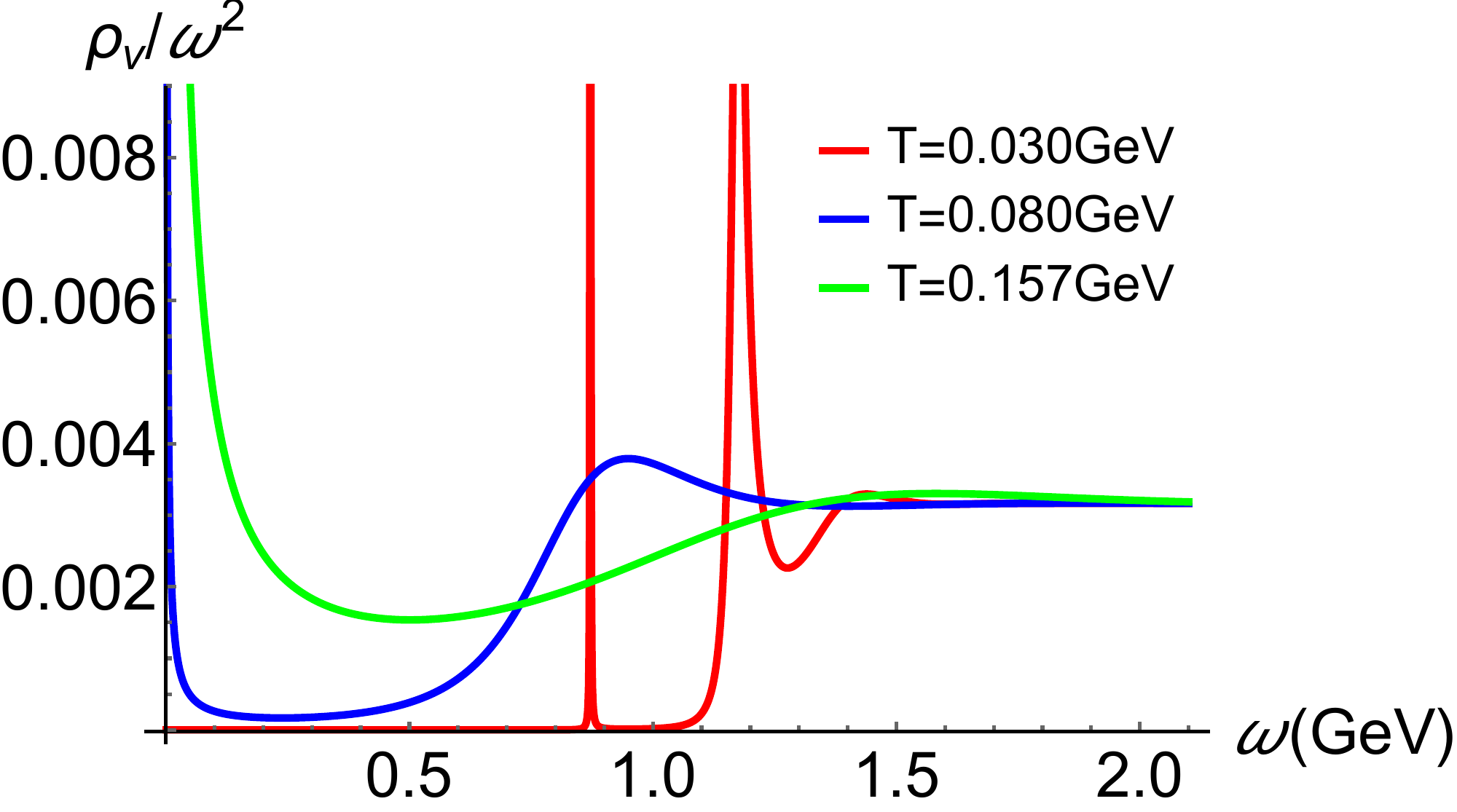}
        \put(80,50){\bf{(a)}}
    \end{overpic}
    \begin{overpic}[width=0.45\textwidth]{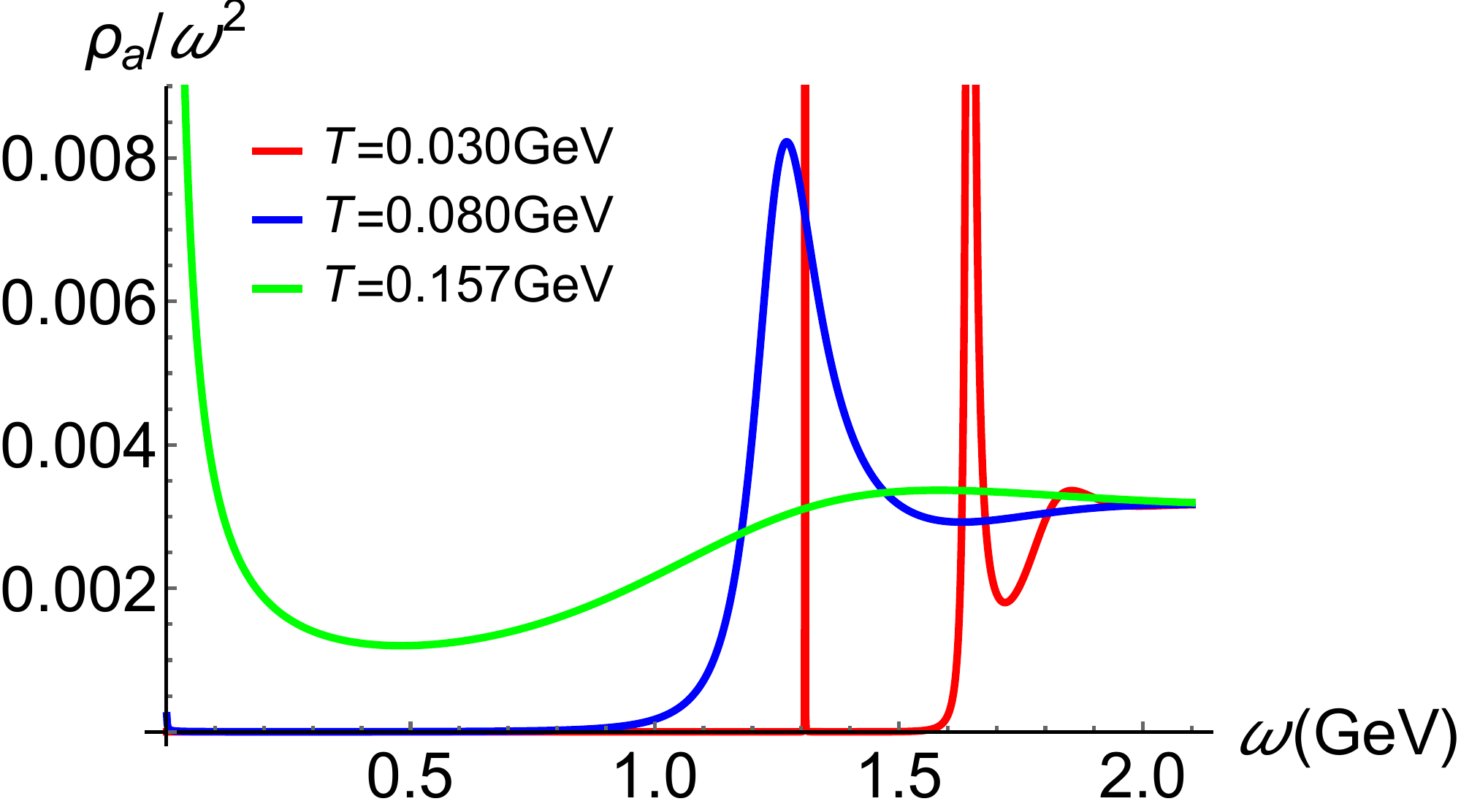}
        \put(80,50){\bf{(b)}}
    \end{overpic}
    \caption{\label{fig:vectorandaxialvectorspectralfunction} The spectral functions of (a) the vector meson and (b) the axial-vector meson at $T=0.030,\ 0.080$ and $0.157$GeV.}
\end{figure}

The spectral function can be obtained from the imaginary part of the retarded Green's function,
\begin{equation}\label{spectralfunction}
\rho(\omega)=-{1\over\pi}{\rm{Im}}[G^R(\omega)].
\end{equation}
Combining Eqs.~\eqref{BC:vectorboundary-omega}, ~\eqref{BC:axialvectorboundary-omega},~\eqref{G:vector-omega},~\eqref{G:axialvector-omega},~\eqref{BC:vectorhorizon-omega}, \eqref{BC:axialvectorhorizon-omega} and~\eqref{spectralfunction}, we can numerically calculate the spectral functions of the vector and the axial-vector mesons.
Figure~\ref{fig:vectorandaxialvectorspectralfunction} shows the spectral functions~\footnote{The spectral functions are rescaled by $\omega^2$, since the spectral functions are proportional to $\omega^2$ in the large $\omega$ limit.} of the vector meson $\rho_{\rm{v}}/\omega^2$ and the axial-vector meson $\rho_{\rm{a}}/\omega^2$ at three different temperatures, $T=0.030,\ 0.080$ and $0.157$ GeV. We find that those peaks arise around the vanishing frequency at relatively high temperature, and the lowest lying states correspond to the smallest non-zero frequency peaks. The peaks of both the vector and axial-vector meson spectral functions shift to the left at low temperature and then to the right side of the $\omega$-axis with the increasing of the temperature, representing the varying of the quasiparticle masses. The quasiparticle masses from the spectral functions are compared with the pole masses from the QNM frequencies in Fig.~\ref{fig:polemass}(c). The black and green dashed lines represent the quasiparticle masses of the vector meson and the axial-vector meson, respectively. They are almost equal to the pole masses from the QNM frequencies at the low temperature. At higher temperature, they increase faster than the pole masses from the QNM frequencies. Nevertheless, the quasiparticle masses of the vector meson and axial-vector meson also degenerate in the chiral symmetric phase.

\subsubsection{Spectral functions versus QNMs}
\begin{figure}[htbp]
    \centering
    \includegraphics[width=.6\textwidth]{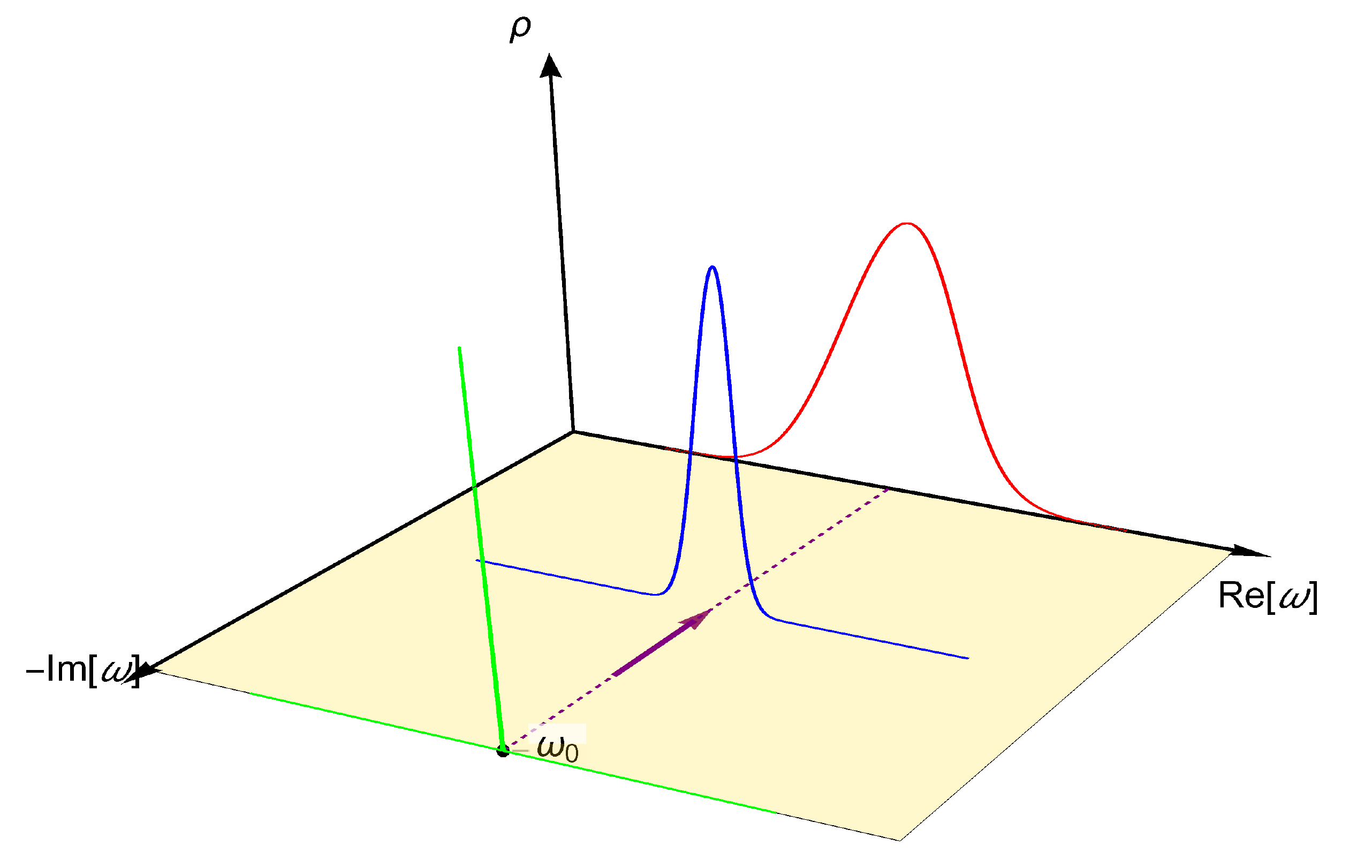}
    \caption{\label{fig:sketch} Schematic diagram to illustrate the connection between the QNM frequency and the spectral function.}
\end{figure}
To clarify how the pole mass and the thermal width (or the QNM frequency)  are connected  with the spectral function, we reconstruct the spectral function with a series of numerical solutions near the complex QNM frequency $\omega_0$. Figure~\ref{fig:sketch} is a cartoon diagram for this approach. Taking the scalar channel as an example, the process can be summarized as three steps: a) Obtain the QNM frequency $\omega_0$ at a particular temperature, such as the black solid point shown in Fig.~\ref{fig:sketch}; b) Calculate a series of numerical solutions in the neighborhood of  $\omega_0$. For example, one can choose frequencies ($\omega^i$) on the line perpendicular to the $\omega_{\rm{Re}}$-axis, as shown with the purple dashed line. One can solve the EOM in Eq.~\eqref{EOM:scalar-omega} with the chosen $\omega^i$ and get a series of coefficients \{$s_1^i$, $s_3^i$\} of the boundary expansions ; c) Fit the spectral function on the complex $\omega$ plane with the calculated data \{$\omega^i$, ${s_1^i}$, ${s_3^i}$\}. The fitting functions for $s_1$ and $s_3$ are
\begin{subequations}
\begin{align}
    s_1(\omega)&=s_{10}+s_{11} (\omega-\omega_0)+s_{12} (\omega-\omega_0)^2+\cdots+s_{1n}(\omega-\omega_0)^n+\cdots,\\
    s_3(\omega)&=s_{30}+s_{31} (\omega-\omega_0)+s_{32} (\omega-\omega_0)^2+\cdots+s_{3n}(\omega-\omega_0)^n+\cdots.
\end{align}
\end{subequations}
Therefore, the fitting spectral function is
\begin{equation}
    \rho_{\rm{FIT}}(\omega)=\frac{4}{\pi}{\rm{Im}}\left[\frac{s_3(\omega)}{s_1(\omega)}\right ].
\end{equation}
Through the fitting function $\rho_{\rm{FIT}}(\omega)$, one can easily get the spectra functions in the neighborhood of $\omega_0$, such as the green, blue and red peaks in Fig.~\ref{fig:sketch}.

\begin{table}[tbp]
\centering
 \begin{tabular}{c|c||c|c}
   \hline
   $s_{10}$ &$2.19 + 0.08 i$&    $s_{30}$ &$ -17.76 + 0.14 i$\\
   $s_{11}$ & $-7.12 + 7.32 i$& $s_{31}$&$46.28 - 62.86 i$\\
   $ s_{12}$&$3.97 - 25.63 i$& $ s_{32}$&$4.82 + 193.76 i$\\
   $s_{13}$& $11.19 +35.42 i$& $s_{33}$&$-131.57 -236.89 i$\\
   $s_{14}$&$-20.56 - 24.48 i$& $s_{34}$ &$181.15 + 141.90 i $\\
   $s_{15}$&$14.44 + 8.37 i$ & $s_{35}$&$-111.92 - 38.60 i$\\
   $ s_{16}$&$-4.76 - 1.06 i$&  $ s_{36}$&$ 33.59 + 1.58 i$\\
   $s_{17}$&$0.61 -0.04 i $ & $s_{37}$&$ -3.96 +0.86 i$\\
 \hline
 \end{tabular}
 \caption{The fitting parameters at $T=0.1$ GeV}\label{tab:fit1}
\vspace{.8cm}
 \begin{tabular}{c|c||c|c}
   \hline
   $s_{10}$ &$0.06$&    $s_{30}$ &$-0.46$\\
   $s_{11}$ & $-0.04 i$& $s_{31}$&$0.01+0.47 i$\\
   $ s_{12}$&$-0.22-0.03 i$& $ s_{32}$&$1.79+0.17$\\
   $s_{13}$& $0.13-0.17 i$& $s_{33}$&$-0.70+1.51 i$\\
   $s_{14}$&$0.14+0.22 i$& $s_{34}$ &$-0.72-1.19 i$\\
   $s_{15}$&$-0.17-0.01 i$ & $s_{35}$&$0.90-0.18 i$\\
   $ s_{16}$&$0.04-0.05 i$&  $ s_{36}$&$-0.26+0.24 i$\\
   $s_{17}$&$ 0.01 i $ & $s_{37}$&$ 0.01-0.02 i$\\
 \hline
 \end{tabular}
 \caption{The fitting parameters at $T=0.153$ GeV}\label{tab:fit2}
\vspace{.8cm}

 \begin{tabular}{c|c||c|c}
   \hline
   $s_{10}$ &$0$&               $s_{30}$ &$0.07$\\
   $s_{11}$ & $-0.20 i$&           $s_{31}$&$1.80 i$\\
   $ s_{12}$&$0.03$&   $ s_{32}$&$0.11$\\
   $s_{13}$& $0.13 i$&    $s_{33}$&$-0.13 i$\\
   $s_{14}$&$-0.09$&    $s_{34}$ &$0.51$\\
   $s_{15}$&$-0.09 i$ &    $s_{35}$&$0.08 i$\\
   $ s_{16}$&$0.05$&    $ s_{36}$&$-0.11$\\
   $s_{17}$&$ 0.01 i $ &          $s_{37}$&$ 0.01 i$\\
 \hline
 \end{tabular}
 \caption{The fitting parameters at $T=0.175$ GeV}\label{tab:fit3}
 \end{table}

\begin{figure}[tbp]
    \centering 
    \begin{overpic}[width=0.45\textwidth]{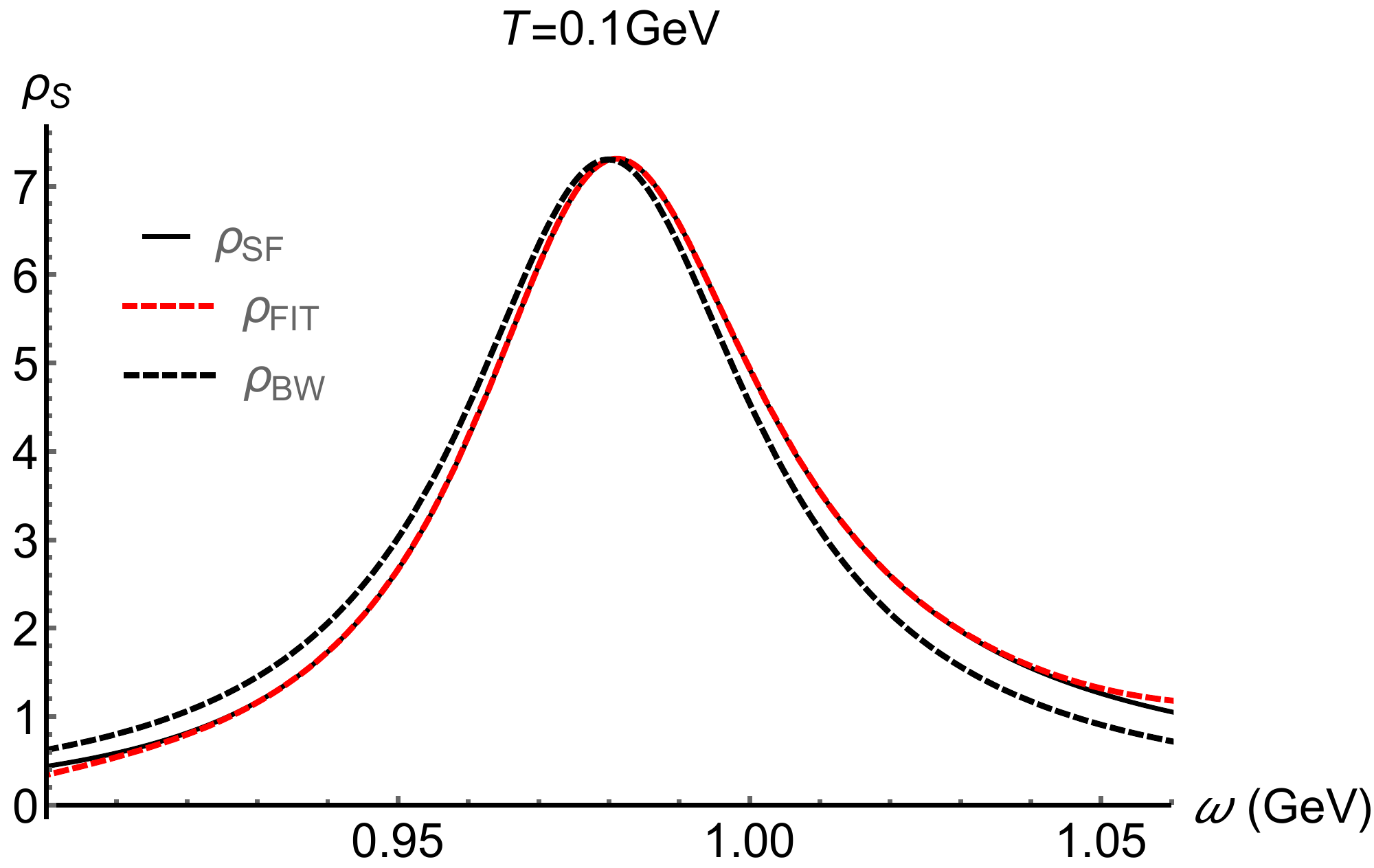}
        \put(73,52){\bf{(a)}}
    \end{overpic}
    \begin{overpic}[width=0.45\textwidth]{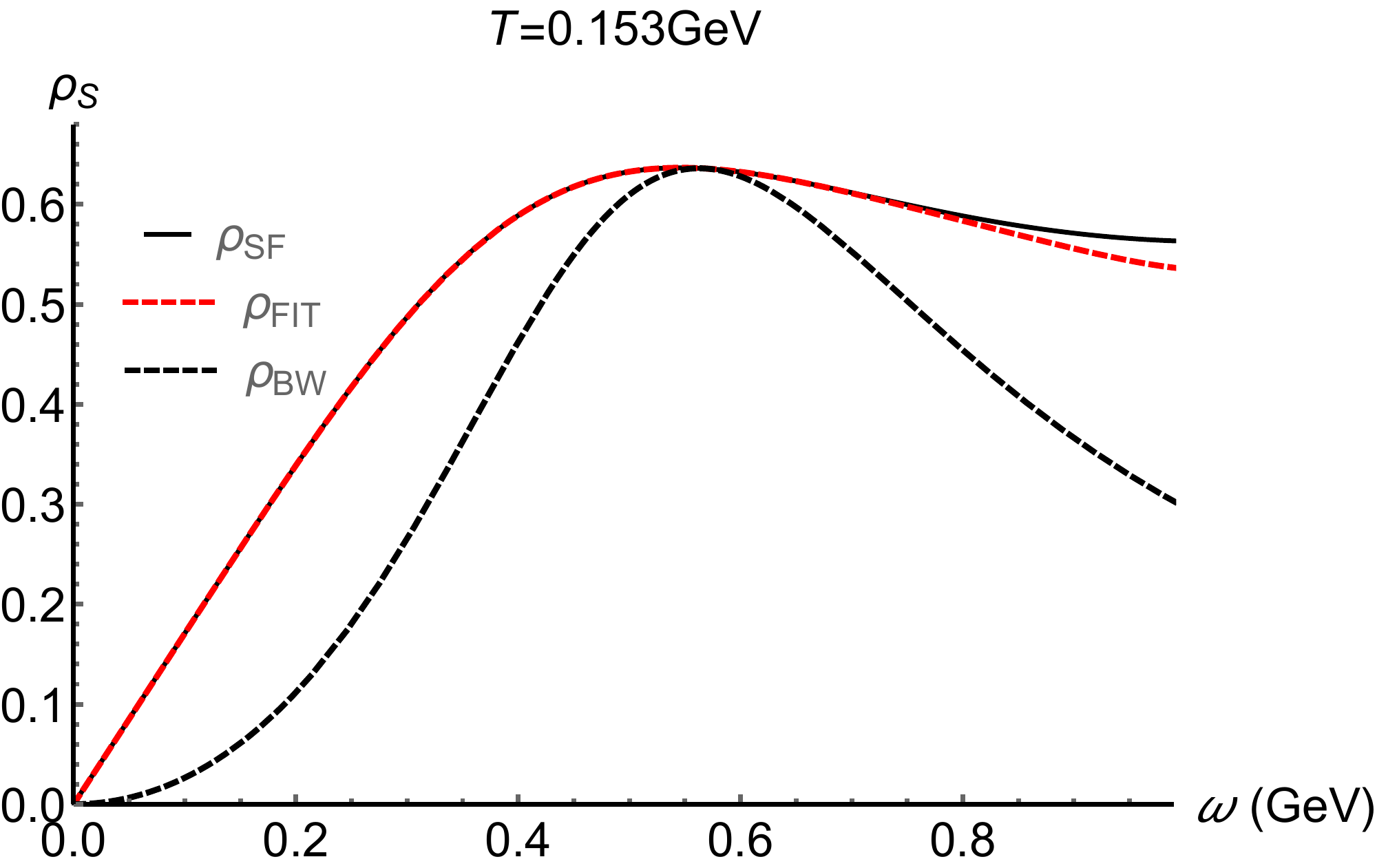}
        \put(73,52){\bf{(b)}}
    \end{overpic}
    \begin{overpic}[width=0.45\textwidth]{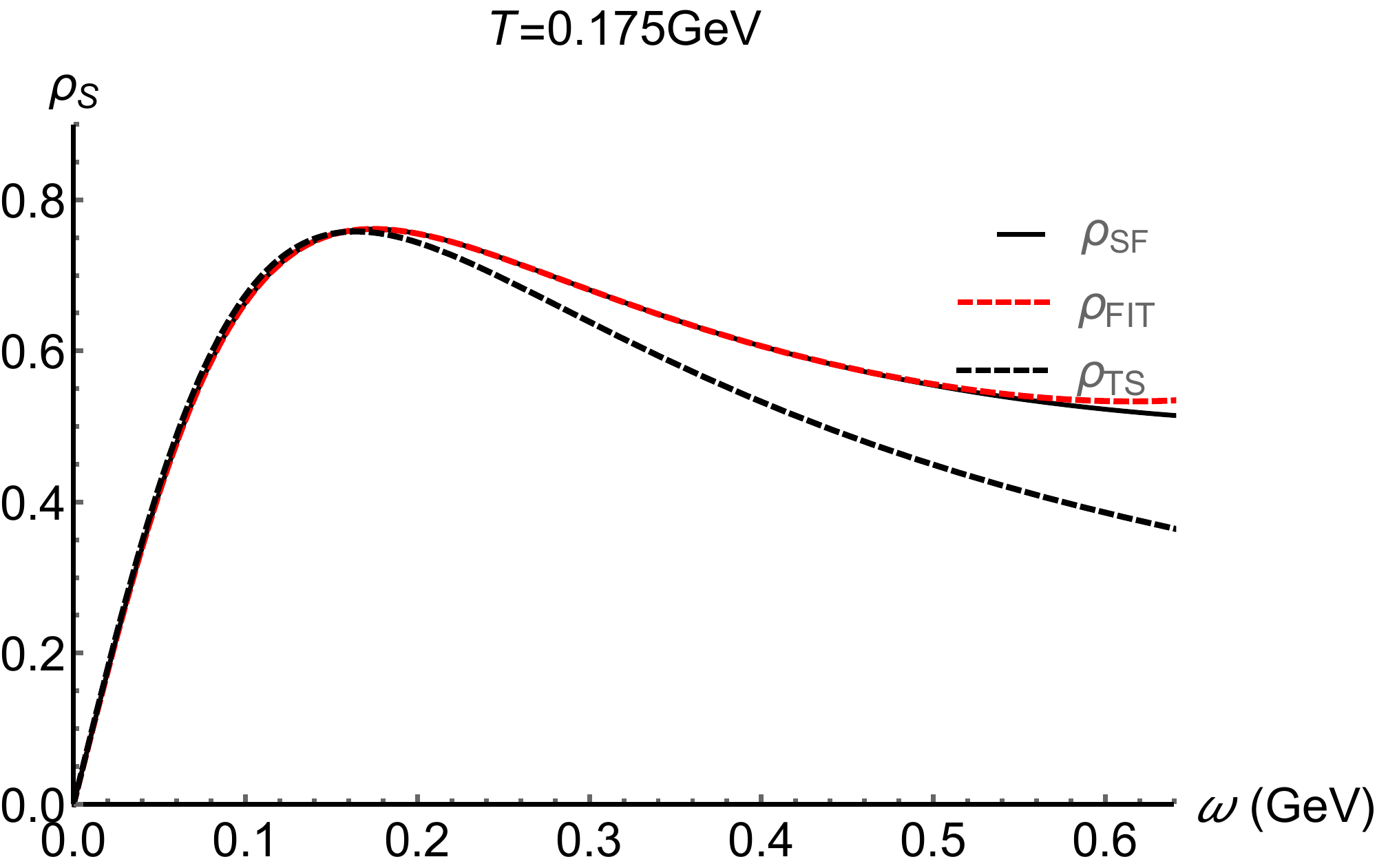}
        \put(73,52){\bf{(c)}}
    \end{overpic}
    \caption{\label{fig:scalar-compare:QNMsp} Comparison among the spectral functions ($\rho_{\rm{SF}}$), the fitted-spectral functions($\rho_{\rm{FIT}}$) and Breit-Wigner formula ($\rho_{\rm{BW}}$) of the scalar meson at (a) $T=0.100$GeV and (b) $T=0.153$GeV, and comparison with $\rho_{\rm{TS}}$ (Eq.~\ref{fit:TS}) of the scalar meson at (c) $T=0.175$GeV.}
\end{figure}

\begin{figure}[thbp]
    \centering
    \begin{overpic}[width=0.47\textwidth]{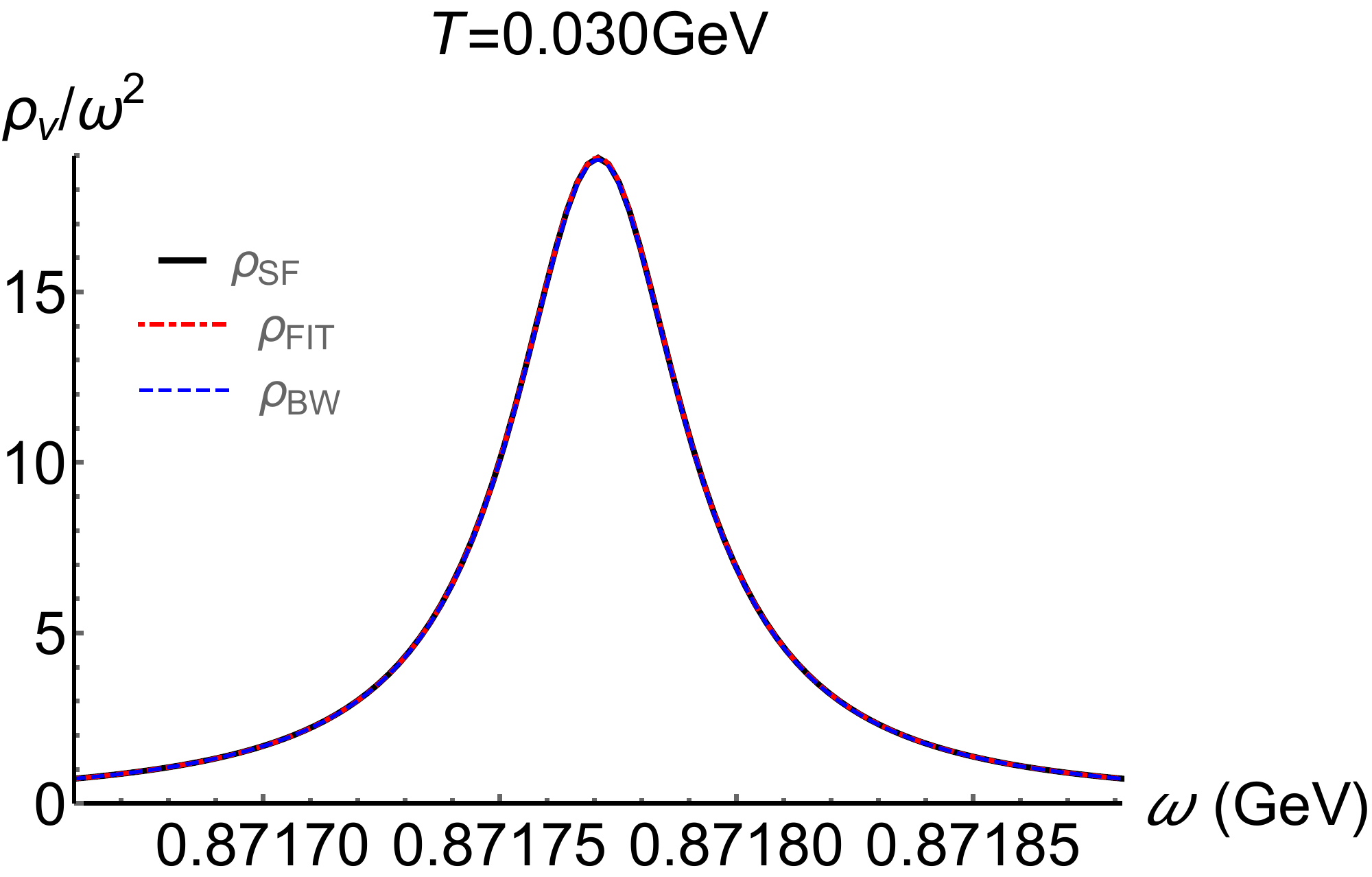}
        \put(80,50){\bf{(a)}}
    \end{overpic}
    \begin{overpic}[width=0.47\textwidth]{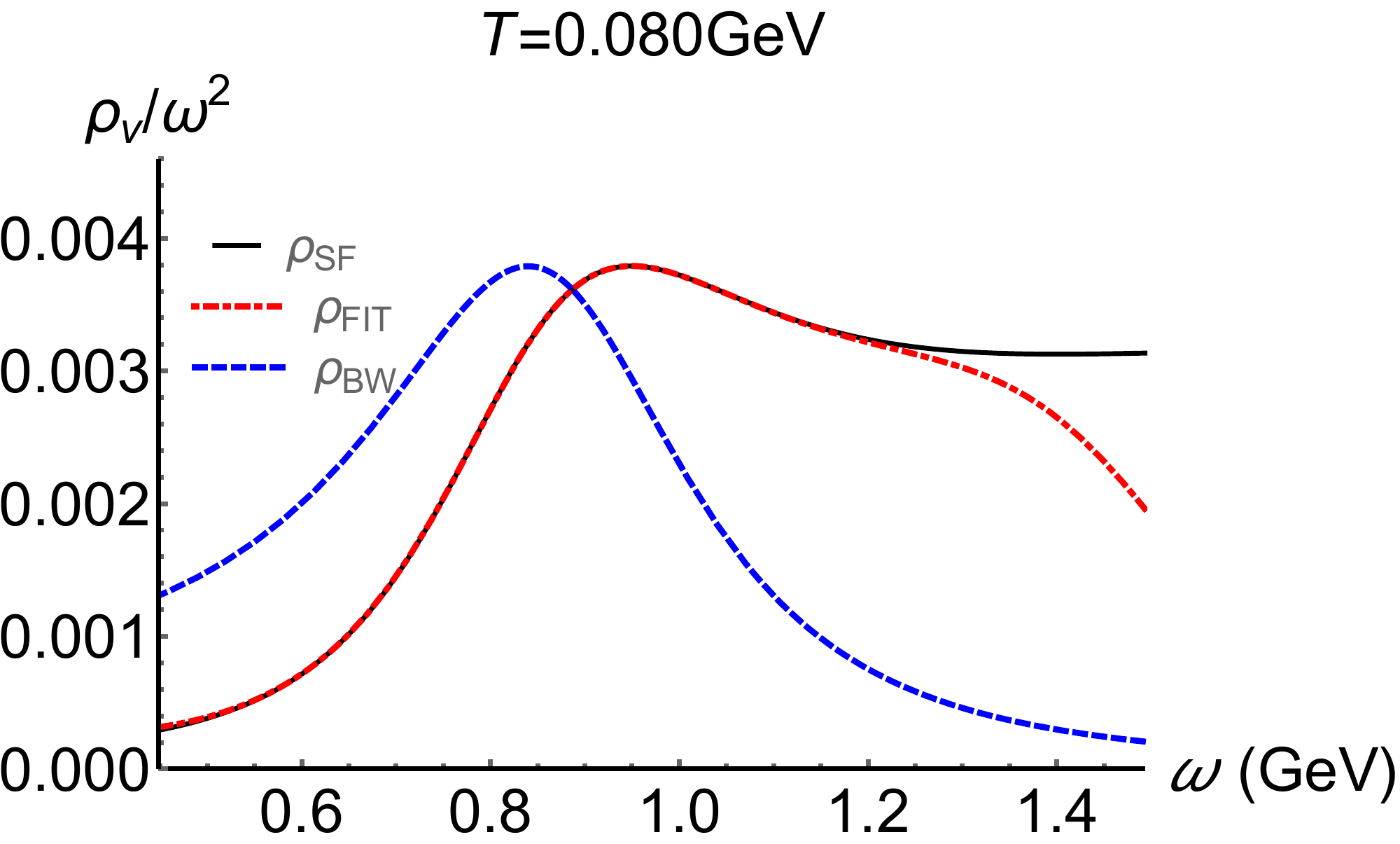}
        \put(80,50){\bf{(b)}}
    \end{overpic}
    \begin{overpic}[width=0.47\textwidth]{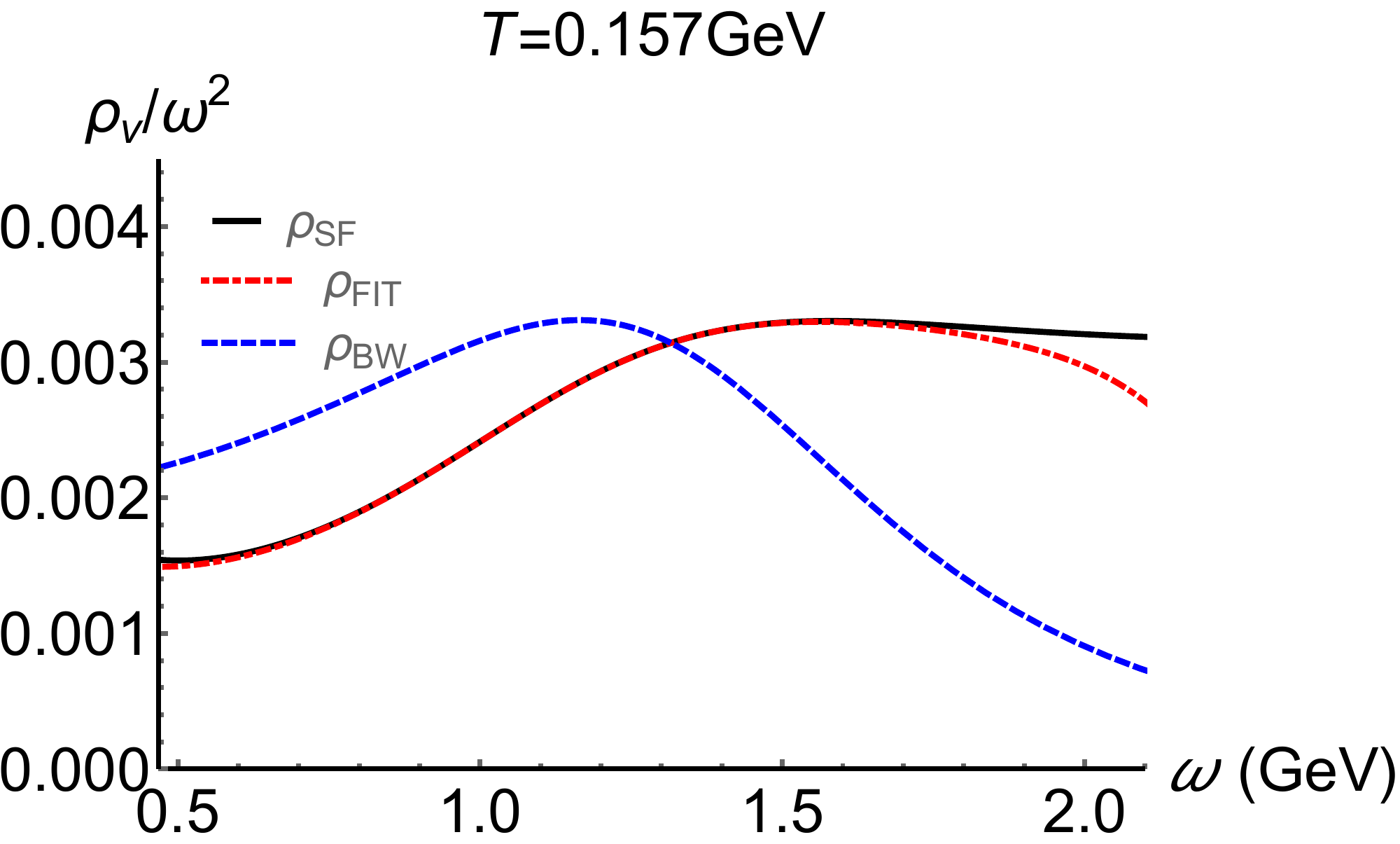}
        \put(80,50){\bf{(c)}}
    \end{overpic}
    \caption{\label{fig:comparevector} Comparison among the spectral functions ($\rho_{\rm{SF}}$), the fitted-spectral functions($\rho_{\rm{FIT}}$) and Breit-Wigner formula ($\rho_{\rm{BW}}$)  of the vector meson at (a) $T=0.030$GeV, (b) $T=0.080$GeV and (c) $T=0.157$GeV.}
\end{figure}

Following the reconstruction steps, in the chiral limit, we get three different spectral functions of the scalar meson on the real $\omega$-axis at  $T=0.1$, $0.153$ and $0.175$ GeV. The fitting parameters are shown in Table.~\ref{tab:fit1}-\ref{tab:fit3}, in which one can find that the expansions are approximately convergent, and the higher order parameters are negligible. Figure~\ref{fig:scalar-compare:QNMsp} shows the spectral functions $\rho_{\rm{FIT}}$ and the comparison with the Breit-Wigner formula $\rho_{\rm{BW}}$ and the spectral functions $\rho_{\rm{SF}}$, which are directly calculated with the imaginary parts of the Retarded Green's functions. The spectral functions $\rho_{\rm{SF}}$ obtained at different temperatures are all well consistent with $\rho_{\rm{FIT}}$. In~Fig.~\ref{fig:scalar-compare:QNMsp}(a), the thermal width $\Gamma_{\rm{S}}/2(T=0.1)=0.026$ GeV is much smaller than the pole mass from the QNM frequency $m_{\rm{S,pole}}(T=0.1)=0.98$ GeV. As a result, the peaks can be well described by the Breit-Wigner formula,
\begin{equation}\label{fit:BW}
\rho_{\rm{BW}}\sim\frac{ m_{\rm{pole}}\Gamma \omega^2}{(\omega^2-m_{\rm{pole}}^2)^2+m_{\rm{pole}}^2\Gamma^2}.
\end{equation}

In the neighborhood of $T_c$, such as~Fig.~\ref{fig:scalar-compare:QNMsp}(c), when the pole are pure imaginary, the thermal width is dominant in the spectral function, then one can simply fit the spectral function with the leading  terms $s_{10}, s_{11}$ and $s_{30}$. The fitting form is
\begin{eqnarray}\label{fit:TS}
    \rho_{\rm{TS}}\sim \frac{\Gamma/2 \omega}{\omega^2+[\Gamma/2]^2}.
\end{eqnarray}
As shown in Fig.~\ref{fig:scalar-compare:QNMsp}(c), the peak of $\rho_{\rm{S,SF}}$ locates at $m_{\rm{S,SF}}\approx \Gamma_{\rm{S}}/2$ and is well consistent with $\rho_{\rm{TS}}$.

As shown in Fig.~\ref{fig:scalar-compare:QNMsp}(b), at $T=0.153$ GeV, the thermal width is comparable to the pole mass from the QNM frequency, $\Gamma_{\rm{S}}/2=0.422$ GeV $\approx $ $m_{\rm{\rm{S,pole}}}=0.445$ GeV. Both the pole mass from the QNM frequency and the thermal width are not negligible. One cannot simply describe the peak with $\rho_{\rm{BW}}$ or $\rho_{\rm{TS}}$.

Since the behaviors of the vector meson are very different from the scalar one, we have also checked the relationships for the vector meson. We fit the peaks of the vector meson spectral functions at $T=0.030,\ 0.080$ and $0.157$ GeV, respectively, as shown in Fig.~\ref{fig:comparevector}. The fitting curves $\rho_{FIT}$ are in perfect agreement with the spectral functions $\rho_{SF}$ at any temperature. However, it is verified again that the Breit-Wigner formula $\rho_{BW}$ is only applicable at low temperature and with small thermal width.

\section{Summary and discussion}\label{V}
In this work, we have investigated the thermal properties of the light mesons, including the screening masses, the pole masses and the thermal widths in the framework of holographic QCD. We have also studied the relationship between the pole masses from the QNM frequencies and the quasiparticle masses from the spectral functions. The nearest QNM frequencies on the complex $\omega$ plane always determine the peaks of the spectral functions.

In the chiral limit, the pole masses, the thermal widths and the screening masses of the scalar meson and the pion all become zero at the chiral critical temperature $T_c$. Below $T_c$, the pion with zero mass and thermal width is the Nambu-Goldstone boson of the chiral phase transition. The pole mass of the scalar meson from the QNM frequency monotonically decreases to zero at $T_{\rm{S,0}}$ ($T_{\rm{S,0}}<T_c$). However, the thermal width increases first and then decreases with the increasing temperature.  Above $T_c$, the masses of the scalar meson and the pion merge and monotonically increase with the increasing temperature. All these behaviors describe the chiral phase transition well at the hadronic spectrum level. We also compare the pole masses with the quasiparticles masses investigated in Ref.~\cite{Cao:2020ryx}. As to the physical quark mass, the chiral phase transition turns to be a crossover. Below the crossover temperature $T_{\rm{cp}}$, the pole mass and screening mass of the pion split with finite temperature, and they are monotonically decrease and increase with the increasing temperature, respectively. The qualitative behaviors of the screening masses are in good agreement with the LQCD simulations in Refs.~\cite{Cheng:2010fe,Bazavov:2019www}. Besides, the quasiparticle masses can be related to the pole masses and the thermal widths from the QNM frequencies with $m_{\rm{SF}}^2\approx m_{pole}^2+(\Gamma/2)^2$. Also, the rapid increase of the thermal widths indicates the dissociations of light mesons  at high temperature.

In the vector and axial-vector channels with physical quark mass $m_q=3.22$ MeV, the pole masses only have small changes at low temperature. Especially, the pole mass of vector meson decreases by about $7.5 \%$, which is in good agreement with the theoretical models as well as the experimental data~\cite{Arnaldi:2006jq,Rapp:2009yu}.  The thermal widths of vector and axial-vector mesons monotonically increase with the increasing temperature. Above $T_{cp}$, the vector and axial-vector mesons degenerate, which can be also considered as a signal for the chiral symmetry restoration.

\appendix
\section{Screening masses in the high-temperature limit}\label{high temperature limit}

\begin{figure}
  \centering
  \includegraphics[width=0.5\textwidth]{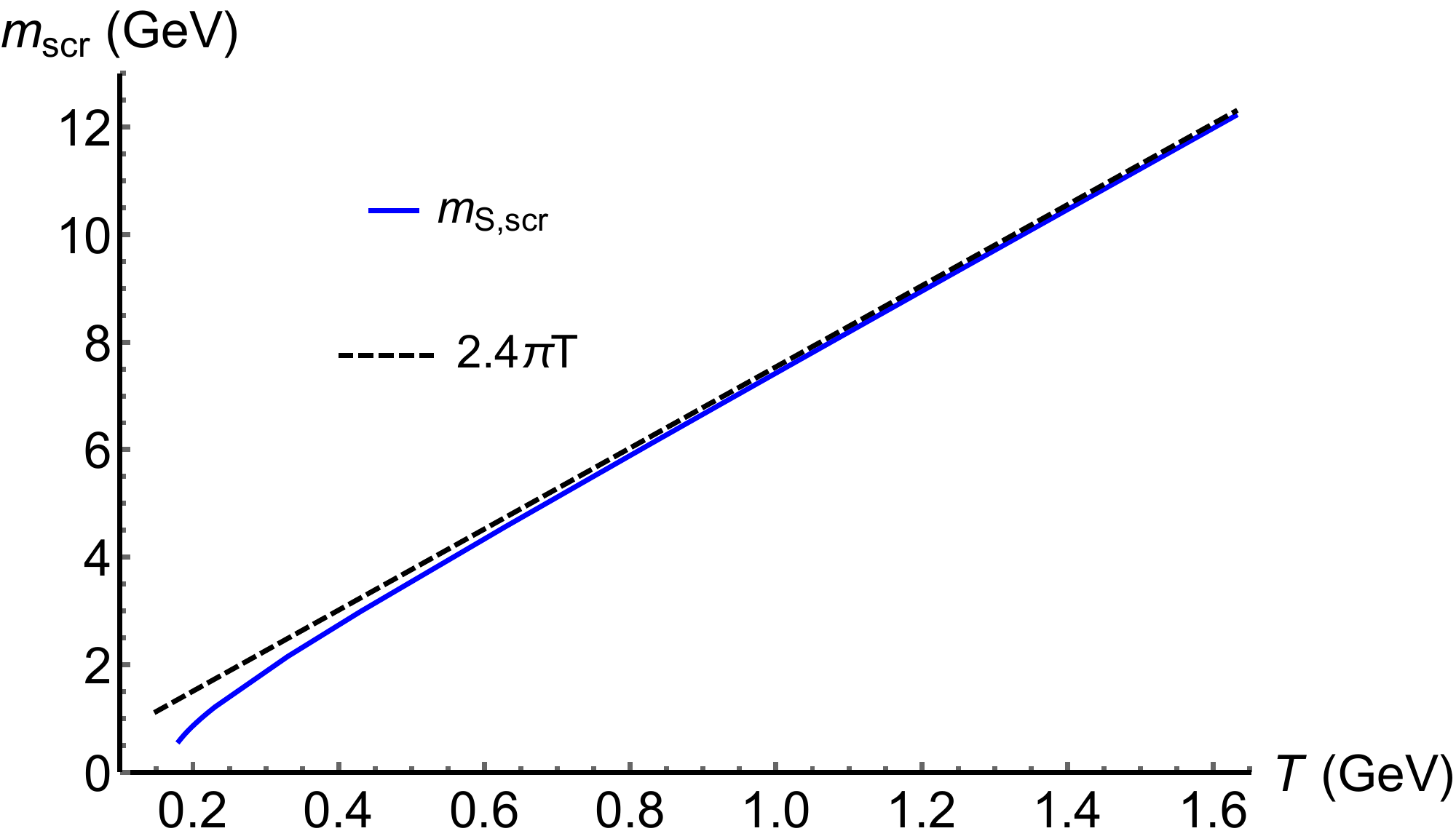}
  \caption{The temperature dependence of the screening mass of the scalar meson in the high temperature limit with $m_q=0$.}\label{checkscalarscr}
\end{figure}

In this section, we will provide a brief proof that the screening mass increases linearly with $T$ at very high temperature. The basic idea to reach this conclusion bases on a simple variable transformation. We take the scalar channel in the chiral limit as an example to show the main strategy. With the variable transformation $t\equiv\frac{z}{z_h}=\pi T z, p_T\equiv\frac{p}{\pi T}$, Eq.~(3.8) becomes
\begin{equation}
\ddot{S}-\left (\frac{3}{t}+\frac{4 t^3}{1-t^4}+\frac{2\mu_g^2 }{\pi^2 T^2}t\right )\dot{S}-\left [\frac{p_T^2}{1-t^4}-\frac{3+\mu_c^2 t^2/(\pi^2T^2)}{(1-t^4)t^2}\right ]S=0,
\end{equation}
in the high temperature limit (since $\chi\equiv0$ above $T_c$ in the chiral limit). Here, the `dot' stands for the derivative with respect to $t$.
The terms with $\mu_g$ and $\mu_c$ would be suppressed by $T$ in the high temperature limit, thus the above equation becomes
\begin{equation}\label{simplifiedscalar}
\ddot{S}-\left (\frac{3}{t}+\frac{4 t^3}{1-t^4}\right )\dot{S}-\left [\frac{p_T^2}{1-t^4}-\frac{3}{(1-t^4)t^2}\right ]S=0.
\end{equation}

The above simplified scalar meson EOM is still a second-order ordinary differential equation with multiple singularities. However, except for the $p_T^2$ term, the other terms in the Eq.~\eqref{simplifiedscalar} do not explicitly depend on temperature. Once we find the specific solution which satisfies the pole condition of the Green's function, we could prove the linear temperature dependence of screening mass in the high temperature limit. We do not have the analytical solution; nevertheless we can numerically solve this EOM. We have the boundary conditions at the UV boundary ($t=0$) as
\begin{eqnarray}
S(t\rightarrow 0)=s_{1,T} t + s_{3,T} t^3-\frac{1}{2}p_T^2 s_{1,T} t^3 \log (t)+\mathcal{O}(t^4),
\end{eqnarray}
and at the horizon ($t=1$) as
\begin{eqnarray}
S(t\rightarrow 1)=s_{h0,T}-\frac{s_{h0,T}}{4 } (p_T^2+3)  (1-t)+\mathcal{O}[(1-t)^2],
\end{eqnarray}
where $s_{1,T}$, $s_{3,T}$ and $s_{h0,T}$ are the integration constants. To extract the screening mass, $s_{1,T}=0$ is required. Combining the EOM and the boundary conditions, we can numerically solve the EOM and obtain
 \begin{equation}
p_T^2=-5.74.
\end{equation}
Thus, we have
\begin{equation}
m_{\rm{S,scr}}=\sqrt{-\pi^2 T^2 p_T^2}\approx 2.40 \pi T.
\end{equation}
Therefore, it is verified that the screening mass is linearly proportional to temperature in the high temperature limit. We also give  a numerical check up to $T=1.63$ GeV in Fig.\ref{checkscalarscr}. It is shown that the screening mass of the scalar meson approach $m_{\rm{S,scr}}\approx 2.40 \pi T$ with the increasing of the temperature.

For a more general situation with a finite quark mass, $m_q\neq 0$, $\chi(z)$ approaches $m_q \zeta  z$ in the high temperature limit, so that the term including $\chi$ in Eq.~\eqref{EOM:scalar} can still be neglected, and the EOM can be simplified to Eq.~\eqref{simplifiedscalar}. In other channels, one can also show the corresponding screening masses vary linearly with temperature in the high temperature limit.

\acknowledgments
We would like to thank the useful discussion with  Yuyiu Lam, Lang Yu and Xinyang Wang. H.L. is supported by the National Natural Science Foundation of China under Grant No. 11405074. D.L. is supported by the National Natural Science Foundation of China under Grant No.11805084, the PhD Start-up Fund of Natural Science Foundation of Guangdong Province under Grant No. 2018030310457 and Guangdong Pearl River Talents Plan under Grant No. 2017GC010480.


\end{document}